\documentclass[%
 amsmath,amssymb,
 aps,
 pre,
 onecolumn,
]{revtex4-2}

\usepackage{graphicx}
\usepackage{dcolumn}
\usepackage{bm}
\usepackage{xcolor}
\usepackage{commath}
\usepackage{hyperref}
\usepackage{cleveref}
\usepackage{sansmath}

\begin{document}

\title{Statistical field theory for a passive vector model with spatially linear advection}

\author{Lukas Bentkamp}
 \email{lukas.bentkamp@uni-bayreuth.de}
\affiliation{Theoretical Physics I, University of Bayreuth, Universit\"atsstr.\ 30, 95447 Bayreuth, Germany}
\author{Michael Wilczek}
 \email{michael.wilczek@uni-bayreuth.de}
\affiliation{Theoretical Physics I, University of Bayreuth, Universit\"atsstr.\ 30, 95447 Bayreuth, Germany}

\date{\today}

\begin{abstract}
One challenge in developing a statistical field theory of turbulence is the analysis of the functional equations that govern the complete statistics of the flow field. Simplified models of turbulence may help to develop such a statistical framework. Here, we consider the advection and stretching of an incompressible passive vector field by a spatially linear stochastic field as a model for small-scale turbulence. The model encompasses non-Gaussian statistics due to an intermittent energy flux from large scales to small scales, thereby displaying hallmark features of turbulence. We explore this model using the Hopf functional formalism, which naturally leads to a decomposition of the complex non-Gaussian statistics into Gaussian sub-ensembles based on different realizations of the advecting field. We then characterize intermittency of the model using a numerical implementation, which takes advantage of this statistical decomposition.
\end{abstract}

\maketitle

Capturing the complex, scale-dependent statistics of fully developed turbulence remains a challenge until today. Statistical approaches that start from the Navier-Stokes equation to derive exact relations for statistical moments or probability density functions (PDFs) are limited due to the closure problem \cite{Monin2013BothVolumes,pope2000,DavidsonPT2005}. The closure problem refers to the fact that exact relations in turbulence always turn out to contain more unknowns than equations relating them. In principle, this is solved by functional approaches such as the ones established by Hopf \cite{hopf1952,Monin2013BothVolumes} or by  Lewis and Kraichnan~\cite{lewis1962}. They contain the statistics of the entire field and thus they are closed, but obtaining tangible results for them remains challenging. Recent reviews of functional approaches to turbulence are provided by Kollmann~\cite{kollmann2019} and Ohkitani~\cite{OhkitaniPTRSMPES2022}. 

A useful alternative to the investigation of the full Navier-Stokes equation is provided by surrogate models, i.e., simplified models that maintain or mimic relevant features of turbulence while allowing for analytical insights. One such model was introduced by Kraichnan \cite{KraichnanTPoF1968}, who studied the evolution of a passive scalar in a turbulent field. As a key simplification, he replaced the turbulent field by a Gaussian field that is delta correlated in time and whose spatial correlations are prescribed. The model, which now bears his name, has evolved into a workhorse of turbulence theory. Despite the simple form of the advecting field, the passive scalars in the Kraichnan model reproduce various important aspects of turbulence, such as the emergence of intermittency with scale-dependent statistics and strongly non-Gaussian small-scale features \cite{KraichnanJFM1974a,KimuraPoFAFD1993,KraichnanPRL1994,ShraimanN2000,FalkovichRMP2001}. 

Passive vector models constitute vectorial extensions of passive scalar models~\cite{AdzhemyanE2001,AntonovPRE2003,ArponenPRE2009,YoshidaPRE2000,AradPRE2001,ArponenPRE2010,AntonovPRE2015}. As such, they bear closer resemblance to the Navier-Stokes equation and enable the investigation of the effect of pressure. The class of passive vector models includes the linearized Navier-Stokes equation~\cite{YoshidaPRE2000,ArponenPRE2010}, the linear pressure model~\cite{AradPRE2001,AdzhemyanE2001,ArponenPRE2009}, and the kinematic approximation of magnetohydrodynamics~\cite{KazantsevSPJ1968,VergassolaPRE1996,AntonovPRE2000,VincenziJoSP2002,HnatichPRE2005,ArponenJSP2007}, with all of these references relying on Gaussian advecting fields with either finite or vanishing correlation time. But even in these simplified models, functional approaches are not easily tractable.

Much effort has gone into analyzing passive scalar and passive vector statistics in the inertial-convective range, i.e., the range in which both advecting and passive fields are rough. At large Schmidt numbers, however, there is an additional range of scales between the viscous and diffusive cut-offs, where a smooth velocity field acts on a rough scalar field, which is called the Batchelor regime~\cite{BatchelorJFM1959,KraichnanTPoF1968}. The dynamics in this range are well approximated by a stochastically fluctuating spatially uniform gradient acting on the passive fields~\cite{KimuraPoFAFD1993,MajdaPF1993,ChertkovPRE1995,BalkovskyPRE1999}, in other words, a spatially linear velocity field. Some field-theoretic approaches have been used to derive passive scalar statistics in the Batchelor regime~\cite{ShraimanPRE1994,FalkovichPRE1996,BernardJoSP1998,GambaJoSP1999,KolokolovJETP1999,SonPRE1999}.

Here, we contribute to these lines of research in the framework of a passive vector model advected by a Gaussian, temporally delta-correlated, and spatially linear velocity field. The latter restriction greatly facilitates the analytical treatment, enabling interesting insights in the context of Hopf's functional approach to turbulence~\cite{hopf1952}. By including advection, stretching, and pressure effects, the model features a transfer of energy across scales, small-scale viscous dissipation and intermittent non-Gaussian fluctuations. At the same time, based on the Hopf equation for the characteristic functional, non-Gaussianity in this model can be understood as the result of heterogeneous stretching across different members of the statistical ensemble, naturally suggesting a description in terms of Gaussian sub-ensembles. The model thereby offers an intuitive understanding of small-scale intermittency. We illustrate the statistical properties of the model by means of numerical simulations that make use of the Gaussian decomposition.

\section{Model equations}

In the following we consider a three-dimensional incompressible vector field $\bm u(\bm x,t)$ on an infinite spatial domain $\bm x \in \mathbb{R}^3$. This field evolves according to
\begin{equation}
  \label{eq:passive_vector_model}
  \partial_t \bm u + \bm v \cdot \nabla \bm u = \gamma \bm u \cdot \nabla \bm v  -\nabla p + \nu \nabla^2 \bm u + \bm F \, .
\end{equation}
Here, $\bm v(\bm x, t)$ is an incompressible ($\nabla \cdot \bm v = 0$), Gaussian, temporally delta-correlated and spatially linear velocity field with isotropic statistics and takes the form
\begin{equation}
  \bm v(\bm x,t) = \bm V(t) + \mathrm{B}(t) \bm x \, .
\end{equation}
Moreover, $\bm V$ is a spatially constant fluctuating mean velocity with isotropic statistics
\begin{align}
  \langle \bm V \rangle &= \bm 0 \\
  \langle V_i(t) V_j(t') \rangle &= \alpha \delta_{ij} \delta(t-t') \, ,
\end{align}
which will be largely unimportant for the following considerations.
$\mathrm{B}$ is a spatially constant gradient of an incompressible field ($B_{ii} = 0$, implied summation) with
\begin{align}
  \left\langle  \mathrm{B} \right\rangle &= 0 \\
  \left\langle B_{ik}(t) B_{jl}(t') \right\rangle &= \beta \left( \delta_{ij}\delta_{kl} - \frac{1}{4} \delta_{ik}\delta_{jl} - \frac{1}{4} \delta_{il}\delta_{jk} \right) \delta(t-t') = D_{ikjl} \delta(t-t')
  \label{eq:b_covariance}
\end{align}
The form of this covariance tensor can be found by enforcing isotropy, incompressibility $B_{ii} = 0$, and the first Betchov constraint $\langle B_{ij} B_{ji} \rangle = 0$~\cite{BetchovJFM1956}. It is chosen to be independent of $\bm V$.
The resulting stochastic field both advects and stretches the passive vector field $\bm u$. We call $\bm v \cdot \nabla \bm u$ the advection term and $\gamma \bm u \cdot \nabla \bm v$ the stretching term, although even the differential advection alone leads to deformation and transfer of energy across scales.

The coefficient $\gamma$ in the stretching term determines the type of passive vector model. The choice $\gamma = -1$ corresponds to the linearized Navier-Stokes equation~\cite{YoshidaPRE2000,ArponenPRE2010}, $\gamma = 1$ corresponds to the kinematic dynamo problem in magnetohydrodynamics~\cite{KazantsevSPJ1968,VergassolaPRE1996,AntonovPRE2000,VincenziJoSP2002,HnatichPRE2005,ArponenJSP2007}, and the case $\gamma = 0$ is sometimes referred to as the `linear pressure model'~\cite{AradPRE2001,AdzhemyanE2001,ArponenPRE2009}. In the main text, we will mostly focus on $\gamma = 0$. The more general system with $-1 \leq \gamma \leq 1$ is analyzed in Appendixes A--G.
The pressure $p(\bm x, t)$ keeps the passive vector field incompressible, $\nabla \cdot \bm u = \bm 0$, while it diffuses with viscosity $\nu$. The forcing $\bm F(\bm x, t)$ is independent of $\bm V$ and $\mathrm{B}$, Gaussian, zero-mean, incompressible ($\nabla \cdot \bm F = 0$), temporally delta correlated and statistically homogeneous and isotropic, with covariance
\begin{align} \label{eq:forcing_covariance}
  \langle F_i(\bm x,t) F_j(\bm x',t') \rangle = R_{ij}(\bm x' - \bm x) \, \delta(t-t') \, .
\end{align}
The combined action of forcing and viscosity enables the possibility of a statistically stationary state. 

For the subsequent treatment, it is beneficial to formulate the equations of motion in Fourier space. Using the Fourier transform pair
\begin{align}
  \label{eq:fourier1}
  \bm u(\bm x,t) &= \int \dif  \bm k \, \hat{\bm u}(\bm k,t) \, e^{ \mathrm i \bm k \cdot \bm x } \\
  \hat{\bm u}(\bm k,t) &= \frac{1}{(2\pi)^3} \int \dif  \bm x \, \bm u(\bm x,t) \, e^{-\mathrm i \bm k \cdot \bm x } \,, \label{eq:fourier2}
\end{align}
one can derive the Fourier representation of the passive vector model \eqref{eq:passive_vector_model},
\begin{equation}
  \label{eq:passive_vector_model_fourier}
  \left[\partial_t + \mathrm i k_l V_l(t) + \nu k^2 \right] \hat{u}_i(\bm k,t)  = P_{ij}(\bm k) \left[ k_l B_{lm}(t) \partial_m \hat{u}_j(\bm k,t)+ \gamma B_{jm}(t) \hat{u}_m(\bm k,t) \right] + \hat{F}_i(\bm k,t)  \,,
\end{equation}
where we are using implied summation and $\partial_m$ denotes the partial derivative with respect to $k_m$.  The derivative in $\bm k$-space is reminiscent of models with linear transport in Fourier space, see Reference~\cite{BeckPRR2024}. As usual, the pressure has been eliminated from this equation by introduction of the projector $P_{ij}(\bm k) = \delta_{ij} - k_i k_j/k^2$. The Fourier transform of the forcing is denoted by $\hat{\bm F}(\bm k, t)$ and the forcing spectrum tensor is given by the isotropic form
\begin{align} \label{eq:forcing_spectrum}
  \psi_{ij}(\bm k) &= \frac{1}{(2\pi)^3} \int\dif\bm r\, R_{ij}(\bm r) e^{-\mathrm i \bm k\cdot\bm r}
  = \frac{Q(k)}{4\pi k^2} P_{ij}(\bm k)
  \,,
\end{align}
where $Q(k)$ is the forcing energy spectrum function. The term with a derivative in wavevector space can be rearranged,
\begin{equation}
  k_l B_{lm}(t) \partial_m \hat{u}_j(\bm k,t) = \partial_m \left[ k_l B_{lm}(t) \hat{u}_j(\bm k,t) \right]\,,
\end{equation}
since $\mathrm{B}$ is traceless.

\section{Statistical description}
To analyze the statistics of the model, we introduce the ensemble average $\langle \dots \rangle$, which refers to an average of \emph{all} random quantities in the problem. In our case, these are the realizations of the advecting velocity field, of the forcing, and potentially also of the initial conditions. 

\subsection{Spectral energy budget}
\label{sec:spectral_energy_budget}

The second-order statistics of the model can be described conveniently using the spectral energy tensor. In general, it can be defined as
\begin{equation} \label{eq:spectral_energy_tensor_general}
  \phi_{ij}(\bm k, \bm k',t) = \langle \hat{u}_i(\bm k,t) \hat{u}^*_j(\bm k',t) \rangle \, ,
\end{equation}  
where the star denotes complex conjugation.
As we show in Appendix~\ref{sec:hopf_symmetries}, single-time statistics of the model are homogeneous and isotropic. The spectral energy tensor therefore takes the form \cite[e.g.]{pope2000}
\begin{equation}
  \label{eq:spectral_energy_tensor}
  \phi_{ij}(\bm k, \bm k',t) = \phi_{ij}(\bm k, t) \delta(\bm k' - \bm k) = \frac{E(k,t)}{4\pi k^2}P_{ij}(\bm k) \delta(\bm k' - \bm k) \,,
\end{equation}
i.e.,~it is fully determined by the energy spectrum function $E(k,t)$. Here, $\phi_{ij}(\bm k, t)$ is the spectral energy tensor for a statistically homogeneous field. In Appendix \ref{sec:energy_budget_derivation}, we derive the evolution equation for the spectral energy budget:
\begin{equation}
  \label{eq:spectral_energy_budget}
  \frac{\partial E(k,t)}{\partial t} = -2 \nu k^2 E(k,t) + Q(k) + \frac{\beta}{4} \gamma (7 \gamma+3) E(k,t) - \frac{\beta}{4} \frac{\partial}{\partial k} \left[ 2 k (1+\gamma) E(k,t) - k^2 \frac{\partial E(k,t)}{\partial k} \right] \, .
\end{equation}
The first two terms on the right-hand side correspond to viscous diffusion and external forcing. The third term is a linear forcing caused by the stretching of the passive vector, which can inject or remove energy if $\gamma \neq 0$. The last term on the right-hand side is the energy transfer which couples scales. We can identify the energy flux
\begin{align}
  J(k) = \frac{\beta}{4} \left[ 2 k (1+\gamma) E(k,t) - k^2 \frac{\partial E(k,t)}{\partial k} \right]\,,
\end{align}
that indicates how much energy flows across scales.
As we discuss in Appendix \ref{sec:spectral_budget_general_gamma}, the cases $\gamma = \pm 1$ do not permit statistically stationary states (related to the dynamo effect well known for $\gamma=1$ and instability of the linearized Navier-Stokes equation for $\gamma=-1$, see also References~\cite{AdzhemyanE2001,VincenziJoSP2002,ArponenPRE2010}), so we focus on the case $\gamma=0$ in the following. Nevertheless, many of the results can be generalized to non-zero $\gamma$, as we show in the Appendixes.

For $\gamma = 0$, the spectral energy budget equation simplifies to
\begin{equation}
  \label{eq:spectral_energy_budget_gamma=0}
  \frac{\partial E(k,t)}{\partial t} = -2 \nu k^2 E(k,t) + Q(k) - \frac{\beta}{4} \frac{\partial}{\partial k} \left[ 2 k E(k,t) - k^2 \frac{\partial E(k,t)}{\partial k} \right] \, .
\end{equation}
The typical setting that we will consider is that the forcing is localized at small wavenumbers and viscosity is small. In Navier-Stokes turbulence, this corresponds to the high Reynolds number case. In this setting, eq.~\eqref{eq:spectral_energy_budget_gamma=0} is amenable to asymptotic analysis of the statistically stationary state. If we consider wavenumbers outside of the forcing range but small enough for the viscous term to be small, the equation reduces approximately to 
\begin{equation}
  \label{eq:spectral_energy_budget_without_forcing_and_viscosity}
  0 \approx \frac{\partial}{\partial k} \left[ 2 k E(k,t) - k^2 \frac{\partial E(k,t)}{\partial k} \right] \,,
\end{equation}
which permits the analytical solutions
\begin{equation}
  E(k) \sim k^2 \quad\text{and}\quad E(k) \sim k^{-1}\, .
\end{equation}
These are selected based on the constraints of energy flow. For the first solution, the energy flux $J(k)$ is zero, for the second solution it is a positive constant (flow toward larger wavenumbers). Since no energy is injected at wavenumbers below the forcing range, the low-wavenumber regime has to assume the solution with zero energy flux, $E(k)\sim k^2$. As a result, the energy from the forcing has to flow toward large wavenumbers, which selects the solution $E(k) \sim k^{-1}$ for scales between the forcing and the viscous scale (corresponding to the inertial range in Navier-Stokes turbulence). The scaling $k^{-1}$ is the same as in the Batchelor regime of passive scalar advection~\cite{BatchelorJFM1959,KraichnanTPoF1968,KraichnanJFM1974a}. 
In the viscous range (this would be called diffusive range for passive scalars), the equation reduces approximately to
\begin{equation}
  \label{eq:large_wavenumber_asymptotics_gamma=0}
  0 \approx -\lambda^2 k^2 E(k,t) + k^2 \frac{\partial^2 E(k,t)}{\partial k^2}
\end{equation}
with the length scale $\lambda = \sqrt{8 \nu / \beta}$ which gives the exponentially decaying solution (compare References~\cite{KraichnanTPoF1968,KraichnanJFM1974a})
\begin{equation}
  \label{eq:exponential_cutoff_gamma=0}
  E(k) \sim \exp \left[ -\lambda k \right] \, .
\end{equation}
The exponentially growing solution can be physically ruled out.

Using the method of Green's functions, a complete solution covering the entire wavenumber range can be constructed.
Assuming a statistically stationary state, the equation for the Green's function based on \eqref{eq:spectral_energy_budget_gamma=0} reads
\begin{align}
  \lambda^2k^2 G(k,k') + \frac{\partial}{\partial k} \left[ 2 k G(k,k') - k^2 \frac{\partial G(k,k')}{\partial k} \right] &= \lambda^2k^2 G(k,k') -\frac{\partial}{\partial k} \left[ k^4 \frac{\partial}{\partial k} \frac{G(k,k')}{k^2} \right] \\
  &= \frac{4}{\beta}  \delta(k-k') \, .  \label{eq:greens_function_equation_gamma=0}
\end{align}
It is solved by
\begin{align}
  G(k,k') &= g_1(k') \left( 1 + \frac{1}{\lambda k} \right) \mathrm{e}^{-\lambda k}  + g_2(k') \left( 1 - \frac{1}{\lambda k} \right) \mathrm{e}^{\lambda k} \\
  &+ \frac{2 \lambda}{\beta} \frac{1}{(\lambda k')^2} \left[ \left( 1 - \frac{1}{\lambda k'} \right) \left( 1 + \frac{1}{\lambda k} \right) \mathrm{e}^{-\lambda (k-k')} - \left( 1 + \frac{1}{\lambda k'} \right) \left( 1 - \frac{1}{\lambda k} \right) \mathrm{e}^{\lambda (k-k')} \right] H(k-k')\,, \nonumber
\end{align}
where $H$ denotes the Heaviside step function and $g_1(k')$ and $g_2(k')$ can be chosen to impose the correct asymptotic behavior of the Green's function. The first two terms correspond to solutions of the homogeneous equation, whereas the Heaviside function ensures that the jump condition originating from the delta function in~\eqref{eq:greens_function_equation_gamma=0} is satisfied. Since the Green's function should vanish both for $k\to 0$ and for $k\to\infty$, we obtain
\begin{align}
  g_1(k') &= g_2(k') = \frac{2 \lambda}{\beta} \frac{1}{(\lambda k')^2} \left( 1 + \frac{1}{\lambda k'} \right) \mathrm{e}^{-\lambda k'}\,.
\end{align}
After a rearrangement of terms the complete Green's function reads
\begin{equation}
  G(k,k') = \frac{4\lambda}{\beta (\lambda k')^2}
  \begin{cases}
    \label{eq:complete_greens_function_gamma=0}
    \left( 1 + \frac{1}{\lambda k} \right) \mathrm{e}^{-\lambda k} \left( \cosh(\lambda k') - \frac{\sinh(\lambda k')}{\lambda k'} \right) & k > k' \\
    \left( \cosh(\lambda k) - \frac{\sinh(\lambda k)}{\lambda k} \right) \left( 1 + \frac{1}{\lambda k'} \right) \mathrm{e}^{-\lambda k'} & k \le k'\,.
   \end{cases}
\end{equation}
For a given forcing spectrum, the energy spectrum is then simply obtained from evaluating
\begin{equation} \label{eq:complete_solution_spectral_energy_budget_gamma=0}
  E(k) = \int_0^\infty \dif k' \, Q(k') G(k,k') \, .
\end{equation}
This result is generalized to $\gamma\neq 0$ in Appendix~\ref{sec:spectral_budget_general_gamma}.

Some asymptotic cases are of particular interest. Taking the inviscid limit (corresponding to $\lambda \rightarrow 0$) of the Green's function yields
\begin{equation}
  \lim_{\lambda \rightarrow 0} G(k,k') = \frac{4}{3 \beta}
  \begin{cases}
    \frac{1}{k}  & k > k' \\
    \frac{k^2}{k'^3}  & k \leq k'
   \end{cases}
\end{equation}
recovering the large-scale and inertial range asymptotic behavior discussed above.

Alternatively, consider $k$ beyond the range where the forcing spectrum is non-zero. In this case, only the first case in~\eqref{eq:complete_greens_function_gamma=0} is relevant,
\begin{equation}
  E(k) = \frac{4\lambda}{\beta} \left( 1 + \frac{1}{\lambda k} \right) \mathrm{e}^{-\lambda k} \int_0^\infty \dif k' \, \left( \frac{\cosh(\lambda k')}{(\lambda k')^2} - \frac{\sinh(\lambda k')}{(\lambda k')^3} \right) Q(k')  \, .
\end{equation}
Furthermore, if viscosity is sufficiently small, $\lambda$ will be small such that the hyperbolic function in the integrand can be approximated over the range in which the forcing is non-zero,
\begin{equation}
  \frac{\cosh(\lambda k')}{(\lambda k')^2} - \frac{\sinh(\lambda k')}{(\lambda k')^3} = \frac{1}{3} + \mathcal{O}\left((\lambda k')^2\right) \,,
\end{equation}
yielding (compare Reference~\cite[eq.~(5.14)]{KraichnanJFM1974a})
\begin{equation} \label{eq:stationary_spectrum_solution_largek_smalllambda}
  E(k) \approx \frac{4\lambda \tilde{\varepsilon}}{3 \beta} \left( 1 + \frac{1}{\lambda k} \right) \mathrm{e}^{-\lambda k}
\end{equation}
with the energy injection rate
\begin{align}
  \tilde{\varepsilon} = \int_0^\infty \dif k' \, Q(k') \,.
\end{align}
Note that this is the homogeneous solution of \eqref{eq:spectral_energy_budget_gamma=0} that interpolates between the inertial-range and dissipation-range behavior. 

\subsection{Statistical field theory}
\label{sec:stat_field_theory}
After exploring the second-order statistics of the system, we want to write down the corresponding statistical field theory based on Hopf's functional description of hydrodynamics~\cite{hopf1952}. The complete instantaneous statistics of the velocity field $\bm u(\bm x,t)$ is contained in the characteristic functional \cite{hopf1952,Monin2013BothVolumes}
\begin{equation}
  \Phi[\bm \theta](t) = \left\langle \exp \left[ \mathrm{i} \int \dif \bm x \, \bm \theta(\bm x) \cdot \bm u(\bm x,t) \right] \right\rangle\,,
\end{equation}
where $\bm \theta(\bm x)$ is a real-space test function. Equivalently, we can consider the characteristic functional for the Fourier coefficients $\hat{\bm u}(\bm k,t)$,
\begin{equation} \label{eq:def_full_cf_fourier}
  \Phi[\bm \theta](t) = \left\langle \exp \left[ \mathrm{i} \int \dif \bm k \, \hat{\bm \theta}(-\bm k) \cdot \hat{\bm u}(\bm k,t) \right] \right\rangle \, .
\end{equation}
Note that to obtain this form, we have used the convention
\begin{equation}
  \hat{\bm \theta}(\bm k) = \int \dif  \bm x \, \mathrm{e}^{-\mathrm i \bm k \cdot \bm x} \bm \theta(\bm x)
\end{equation}
for the Fourier transform of the test function which differs from the one for the passive vector field, \eqref{eq:fourier1} and \eqref{eq:fourier2} (as done similarly in~\cite[eqs.~(23) and (44)]{hopf1952} and \cite[vol.~2, eqs.~(11.52) and (28.20)]{Monin2013BothVolumes}).

The evolution equation for the characteristic functional~\eqref{eq:def_full_cf_fourier} can be obtained by taking its time derivative \cite{hopf1952}, inserting Eq.~\eqref{eq:passive_vector_model_fourier} and evaluating the averages term by term. This is detailed in Appendix \ref{app:characteristic_functional_equation}. The resulting Hopf equation (for $\gamma=0$) reads
\begin{align} \label{eq:full_hopf_equation_gamma=0}
  \partial_t \Phi[\bm \theta](t) = \int \dif \bm k \, \hat{\theta}_i(-\bm k) \Biggl(
  &- \nu k^2 \frac{\delta}{\delta \hat{\theta}_i(-\bm k)} - \frac{1}{2} \psi_{ij}(\bm k) \, \hat{\theta}_j(\bm k)
  -\frac{\alpha}{2} k_l \frac{\delta}{\delta \hat{\theta}_i(-\bm k)} \int \dif \bm k'\, \hat{\theta}_m(-\bm k') \, k'_l \, \frac{\delta}{\delta \hat{\theta}_m(-\bm k')}\\
  &+ \frac{1}{2} D_{lmab} \int \dif \bm k' \, \hat{\theta}_c(-\bm k') P_{ij}(\bm k) k_l \partial_m \frac{\delta}{\delta \hat{\theta}_j(-\bm k)}  P_{cd}(\bm k')  k'_a \partial'_b \frac{\delta}{\delta \hat{\theta}_d(-\bm k')} \Biggr) \Phi[\bm \theta](t) \, . \nonumber
\end{align}
This equation displays various structural similarities with the Hopf equation of the full Navier-Stokes equation~\cite{hopf1952,novikov1965,Monin2013BothVolumes}. The first two terms, the viscous and forcing contributions, are identical. While the term related to advection is different, it features second-order functional derivatives similar to the inertial term from the Navier-Stokes equation. In the following, we will see that like the inertial term in Navier-Stokes, the advection term also produces intermittency.

For the following investigation, it is crucial to differentiate between different types of ensemble averages. While $\langle \dots \rangle$ averages over all random quantities, it will turn out to be insightful to just perform the average over the forcing and potentially the initial conditions, denoted by $\langle\dots\rangle_{\bm F}$, while keeping the realization of the advecting velocity fixed. This can be understood as an average conditional on the full time series of $\bm v(\bm x, t)$. The complementary average over the realizations of the advecting velocity field will be denoted by $\langle\dots\rangle_{\bm v}$, with $\langle \dots \rangle = \langle \langle\dots\rangle_{\bm F}\rangle_{\bm v}$. Using this notation, we can introduce the forcing-averaged characteristic functional
\begin{align} \label{eq:def_cond_cf_fourier}
  \Phi^{\bm v}[\bm \theta](t) = \left\langle \exp \left[ \mathrm{i} \int \dif \bm k \, \hat{\bm \theta}(-\bm k) \cdot \hat{\bm u}(\bm k,t) \right] \right\rangle_{\!\! \bm F}\,,
\end{align}
i.e., the functional conditional on a realization of the advecting velocity field $\bm v(\bm x, t)$, which is related to the fully averaged characteristic functional by
\begin{align} \label{eq:mixture_of_subensembles}
\Phi[\bm \theta](t) = \left\langle \Phi^{\bm v}[\bm \theta](t) \right\rangle_{\bm v} \,.
\end{align}

In the same way as for the full characteristic functional, we can derive a Hopf equation for the forcing-averaged characteristic functional. This is detailed in Appendix~\ref{app:characteristic_functional_equation}. The resulting Hopf equation (for $\gamma=0$) reads
\begin{align} \label{eq:conditional_hopf_equation_gamma=0}
  \partial_t \Phi^{\bm v}[\bm \theta](t) = \int \dif \bm k \, \hat{\theta}_i(-\bm k) \Biggl(
  &\left[-\mathrm{i}k_l V_l(t) - \nu k^2\right] \frac{\delta}{\delta \hat{\theta}_i(-\bm k)} - \frac{1}{2} \psi_{ij}(\bm k) \, \hat{\theta}_j(\bm k)+ P_{ij}(\bm k)  k_l B_{lm}(t) \partial_m \frac{\delta}{\delta \hat{\theta}_j(-\bm k)} \Biggr) \Phi^{\bm v}[\bm \theta](t) \, . 
\end{align}
Notably, in this case the advection term contains only a first-order functional derivative. In fact, we can show that this equation is solved by a Gaussian characteristic functional~\cite[p.~119]{hopf1952},
\begin{align} \label{eq:gauss_char_func}
  \Phi_G^{\bm v}[\bm\theta](t) &= \exp\left[-\frac{1}{2} \int\dif\bm k\, \hat{\theta}_i(-\bm k) \phi^{\bm v}_{ij}(\bm k, t)\hat{\theta}_j(\bm k)\right],
\end{align}
where $\phi^{\bm v}_{ij}$ is the conditional spectral energy tensor. In the ansatz, we can assume that conditional single-time statistics are homogeneous (as worked out in Appendix~\ref{sec:hopf_symmetries}), because the differential advection caused by the spatially linear field $\bm v(\bm x, t)$ is uniform in space. We do not assume conditional statistics to be isotropic, due to the dependence on the specific realization of $\bm v(\bm x, t)$. To show that the Gaussian characteristic functional is a solution to the conditional Hopf equation, we insert it into~\eqref{eq:conditional_hopf_equation_gamma=0} and divide out $\Phi_G^{\bm v}$. Then we have
\begin{align} \label{eq:gaussian_into_cond_hopf}
  -\frac{1}{2} \int\dif\bm k\, \hat{\theta}_i(-\bm k) \partial_t \phi^{\bm v}_{ij}(\bm k, t)\hat{\theta}_j(\bm k)
  &= -\int\dif\bm k\, \hat{\theta}_i(-\bm k) \left(\left[-\mathrm{i}k_l V_l(t) - \nu k^2\right] \phi^{\bm v}_{ij}(\bm k, t) + \frac{1}{2} \psi_{ij}(\bm k) \right)\hat{\theta}_j(\bm k) \\
  &\quad - \int\dif\bm k\, \hat{\theta}_i(-\bm k)  P_{ip}(\bm k)  k_l B_{lm}(t) \partial_m \left(\phi^{\bm v}_{pj}(\bm k, t)\hat{\theta}_j(\bm k) \right) \nonumber
\end{align}
Here, we used $\phi^{\bm v}_{ij}(\bm k, t) = \phi^{\bm v}_{ji}(-\bm k, t)$. For the first term with $\bm V(t)$, we find that it is zero by symmetry of the spectrum tensor and anti-symmetry of the coefficient. This is because $\bm V$ performs random sweeping, which does not affect single-time statistics. For the last term, we have
\begin{align}
  &\quad \int\dif\bm k\, \hat{\theta}_i(-\bm k)  P_{ip}(\bm k)  k_l B_{lm}(t) \partial_m \left(\phi^{\bm v}_{pj}(\bm k, t)\hat{\theta}_j(\bm k) \right) \nonumber \\
  &= \int\dif\bm k\, \hat{\theta}_i(-\bm k)  P_{ip}(\bm k)  k_l B_{lm}(t) \partial_m \left(P_{jq}(\bm k)\phi^{\bm v}_{pq}(\bm k, t)\hat{\theta}_j(\bm k) \right) \label{eq:test_func_expr_A}\\
  &= \int\dif\bm k\, \hat{\theta}_i(-\bm k)  P_{ip}(\bm k) P_{jq}(\bm k)  k_l B_{lm}(t) \left(\partial_m \phi^{\bm v}_{pq}(\bm k, t) \right)\hat{\theta}_j(\bm k) \\
  &\quad +\int\dif\bm k\, \hat{\theta}_i(-\bm k)  P_{ip}(\bm k)  k_l B_{lm}(t) \phi^{\bm v}_{pq}(\bm k, t) \partial_m \left(P_{jq}(\bm k)\hat{\theta}_j(\bm k) \right) \nonumber \\
  &= \int\dif\bm k\, \hat{\theta}_i(-\bm k)  P_{ip}(\bm k) P_{jq}(\bm k)  k_l B_{lm}(t) \left(\partial_m \phi^{\bm v}_{pq}(\bm k, t) \right)\hat{\theta}_j(\bm k) \label{eq:test_func_expr_AminusB}\\
  &\quad -\int\dif\bm k\,  \partial_m \left(\hat{\theta}_i(-\bm k)  P_{ip}(\bm k)  \phi^{\bm v}_{pq}(\bm k, t) \right) k_l B_{lm}(t) P_{jq}(\bm k)\hat{\theta}_j(\bm k) \nonumber \\
  &= \frac12 \int\dif\bm k\, \hat{\theta}_i(-\bm k)  P_{ip}(\bm k) P_{jq}(\bm k)  k_l B_{lm}(t) \left(\partial_m \phi^{\bm v}_{pq}(\bm k, t) \right)\hat{\theta}_j(\bm k) \,, \label{eq:test_func_expr_Bhalf}
\end{align}
where we used, first, incompressibility of the conditional spectrum tensor ($P_{jq}(\bm k)\phi^{\bm v}_{pq}(\bm k, t) = \phi^{\bm v}_{pj}(\bm k, t)$), second, a product rule, third, integration by parts on the second term (with $B_{ll} = 0$). In the last step, we realize that the term subtracted in~\eqref{eq:test_func_expr_AminusB} is identical with~\eqref{eq:test_func_expr_A} by simultaneous substitution of $\bm k \to -\bm k$, $i \to j$, $p \to q$ and by the symmetry $\phi^{\bm v}_{pq}(\bm k, t) = \phi^{\bm v}_{qp}(-\bm k, t)$. Then, rearranging the terms leads to the factor $1/2$ in~\eqref{eq:test_func_expr_Bhalf}.

Finally, by generality of $\hat{\bm \theta}(\bm k)$, we can isolate the equation of integrands:
\begin{align} \label{eq:subensemble_spectrum_evo}
  \partial_t \phi^{\bm v}_{ij}(\bm k, t) &= -2 \nu k^2 \phi^{\bm v}_{ij}(\bm k, t) + P_{ip}(\bm k) P_{jq}(\bm k) k_l B_{lm}(t) \partial_m \phi^{\bm v}_{pq}(\bm k, t) + \psi_{ij}(\bm k) \,.
\end{align}
This partial differential equation determines the evolution of the spectral energy tensor $\phi^{\bm v}_{ij}(\bm k, t)$. 
By construction, any solution to~\eqref{eq:subensemble_spectrum_evo} corresponds to a solution of the conditional Hopf equation~\eqref{eq:conditional_hopf_equation_gamma=0} through the ansatz~\eqref{eq:gauss_char_func}. This means that the statistics of the system are described exactly by a mixture of Gaussian sub-ensembles. The mixing can be written as~\eqref{eq:mixture_of_subensembles}, with each sub-ensemble characteristic functional corresponding to a Gaussian field.

This result entails a significant simplification of the statistical field theory of the model (compare Reference~\cite{KimuraPoFAFD1993}, section~V). The single-time field statistics of $\bm u(\bm x, t)$ are comprehensively described by the characteristic functional $\Phi[\bm \theta](t)$, which, given suitable initial conditions, can be written as the superposition of Gaussian characteristic functionals~\eqref{eq:mixture_of_subensembles}. This superposition solves the more complex Hopf equation of the full model~\eqref{eq:full_hopf_equation_gamma=0}. In order to determine the spectral energy tensor of the Gaussian fields, we need to solve~\eqref{eq:subensemble_spectrum_evo}, which can be interpreted as a stochastic partial differential equation (SPDE) due to stochasticity of $\mathrm{B}(t)$, while the stochasticity of the forcing has been removed by averaging, and the sweeping contribution has vanished due to conservation of (single-time) statistical homogeneity. We find that while being amenable to this simplified description, the model still displays intermittency, i.e., an increasing degree of non-Gaussian fluctuations toward the small scales, which originate from the mixture of Gaussian statistics. 
This also holds for $\gamma \neq 0$. The analogous computation is presented in Appendix~\ref{sec:gaussian_solution_conditional_hopf_general_gamma}.
In the remainder of the paper, we explore the statistics of the model in the case $\gamma=0$ numerically by making use of the above structure, solving the SPDE~\eqref{eq:subensemble_spectrum_evo} numerically.

To this end, we mention one more transformation that can be applied to~\eqref{eq:subensemble_spectrum_evo} specifically in the case $\gamma = 0$ (this part does not generalize to general values of $\gamma$, as also mentioned in Appendix~\ref{sec:numerics_general_gamma}). In this case, the tensorial structure of the spectrum tensor $\phi_{ij}(\bm k)$ evolves separately from its amplitude. In fact, given a suitable initial condition and forcing spectrum, the tensorial part of it can be fully absorbed into the product ansatz
\begin{align} \label{eq:spectrum_tensor_ansatz}
  \phi^{\bm v}_{ij}(\bm k, t) &=\left(\delta_{ij} - \frac{k_i k_j}{k^2} \right) \phi^{\bm v}(\bm k, t) 
  = P_{ij}(\bm k) \phi^{\bm v}(\bm k, t)\,.
\end{align}
For this to hold, necessarily the initial condition must be of the form~\eqref{eq:spectrum_tensor_ansatz}.
Inserting this into~\eqref{eq:subensemble_spectrum_evo} yields
\begin{align}
  P_{ij}(\bm k) \partial_t \phi^{\bm v}(\bm k, t) 
  &= -2P_{ij}(\bm k) \nu k^2 \phi^{\bm v}(\bm k, t) + P_{ip}(\bm k) P_{jq}(\bm k) k_l B_{lm}(t) \partial_m \left(P_{pq}(\bm k) \phi^{\bm v}(\bm k, t)\right) + \psi_{ij}(\bm k) \\
  &= -2P_{ij}(\bm k) \nu k^2 \phi^{\bm v}(\bm k, t) + P_{ij}(\bm k) k_l B_{lm}(t) \partial_m \phi^{\bm v}(\bm k, t) + \psi_{ij}(\bm k)\,,
\end{align}
where we used
\begin{align}
  P_{ip}(\bm k) P_{jq}(\bm k) \partial_m P_{pq}(\bm k)
  &= P_{ip}(\bm k) P_{jq}(\bm k) \left(-\frac{\delta_{mp} k_q}{k^2}-\frac{\delta_{mq} k_p}{k^2}+2\frac{k_m k_p k_q}{k^4}\right) \\
  &= 0
\end{align}
due to $P_{ip}(\bm k) k_p = 0$ and the projection property $P_{ip}(\bm k) P_{jq}(\bm k) P_{pq}(\bm k) = P_{ij}(\bm k)$. Using isotropy of the forcing~\eqref{eq:forcing_spectrum}, we have therefore reduced \eqref{eq:subensemble_spectrum_evo} to the simpler, scalar equation
\begin{align} \label{eq:scalar_spec_tensor_evo}
  \partial_t \phi^{\bm v}(\bm k, t) &= - 2\nu k^2 \phi^{\bm v}(\bm k, t) +  k_l B_{lm}(t) \partial_m \phi^{\bm v}(\bm k, t) + \frac{Q(k)}{4\pi k^2}\,.
\end{align}
Given a solution of this scalar equation, we can automatically construct a solution to~\eqref{eq:subensemble_spectrum_evo} by using~\eqref{eq:spectrum_tensor_ansatz}.

\section{Simulations}

\subsection{Numerical implementation}
Thanks to the decomposition property into Gaussian fields, the full (single-time) field statistics of the solution to~\eqref{eq:passive_vector_model} can be inferred from statistics of the conditional spectrum tensor, which in the case $\gamma=0$ satisfies~\eqref{eq:subensemble_spectrum_evo}. Equation~\eqref{eq:subensemble_spectrum_evo} is a stochastic partial differential equation (SPDE) due to stochasticity of the gradient tensor $\mathrm{B}$. The forcing term, however, is non-random and constant in time. Instead of simulating the full model equation~\eqref{eq:passive_vector_model}, we simulate equation~\eqref{eq:subensemble_spectrum_evo} (or even~\eqref{eq:scalar_spec_tensor_evo}), from which we can infer equivalent statistical information about the model. In the following, we describe the details of our numerical implementation. For simplicity, we restrict our presentation to the case $\gamma=0$. The numerics of the general case are discussed in Appendices~\ref{sec:numerics_general_gamma} and \ref{sec:numerical_results_small_gamma}.

 By definition, we know that the spectrum tensor is Hermitian, $\phi^{\bm v}_{ij}(\bm k, t) = {\phi^{\bm v}_{ji}}^\ast(\bm k, t)$, and needs to satisfy $\phi^{\bm v}_{ij}(\bm k, t) = \phi^{\bm v}_{ji}(-\bm k, t)$ in order to correspond to a real-valued field. On the level of~\eqref{eq:subensemble_spectrum_evo}, we observe that all coefficients in the equation are real. So given a real initial condition, we know that the solution $\phi^{\bm v}_{ij}(\bm k, t)$ will remain real for all times $t$. In that case, the spectrum is a real symmetric matrix and symmetric in $\bm k$, i.e.,
\begin{align}
  \phi^{\bm v}_{ij}(\bm k, t) = \phi^{\bm v}_{ji}(\bm k, t) = \phi^{\bm v}_{ij}(-\bm k, t)\,.
\end{align}

\subsubsection{Method of characteristics}
In order to deal with the $\bm k$-derivative in~\eqref{eq:subensemble_spectrum_evo}, we employ the method of characteristics. To this end, define the deformation tensor $\mathrm{W}(t)$ as the solution of
\begin{align} \label{eq:evo_deform_tensor}
  \dod{}{t} W_{ij}(t) &= - W_{il}(t) B_{lj}(t), \quad W_{ij}(0) = \delta_{ij}.
\end{align}
The wavevector characteristics are defined as $\tilde{\bm k}(\bm q, t) = \mathrm{W}(t)^T \bm q$, where $\bm q$ is an initial wavevector. They satisfy the equation~\cite[p.~408]{pope2000}
\begin{align} \label{eq:characteristic_evo}
  \partial_t \tilde{k}_i(\bm q, t) &= - B_{ji}(t) \tilde{k}_j(\bm q, t) \,.
\end{align}
The spectrum along the characteristics,
\begin{align}
  \tilde{\phi}^{\bm v}_{ij}(\bm q, t) &= \phi^{\bm v}_{ij}(\tilde{\bm k}(\bm q, t), t)\,,
\end{align}
follows the equation
\begin{align}
  \partial_t \tilde{\phi}^{\bm v}_{ij}(\bm q, t) &= \partial_t \phi^{\bm v}_{ij}(\tilde{\bm k}, t) + \dpd{\tilde{k}_m}{t} \dpd{}{\tilde{k}_m} \phi^{\bm v}_{ij}(\tilde{\bm k}, t) \\
  &= \partial_t \phi^{\bm v}_{ij}(\tilde{\bm k}, t) - \tilde{k}_l B_{lm}(t) \dpd{}{\tilde{k}_m} \phi^{\bm v}_{ij}(\tilde{\bm k}, t) \\
  &= - 2 \nu \tilde{k}^2 \tilde{\phi}^{\bm v}_{ij}(\bm q, t)
  + \frac{\tilde{k}_i \tilde{k}_l}{\tilde{k}^2} B_{lm}(t) \tilde{\phi}^{\bm v}_{mj}(\bm q, t)
  + \frac{\tilde{k}_j \tilde{k}_l}{\tilde{k}^2} B_{lm}(t) \tilde{\phi}^{\bm v}_{im}(\bm q, t)
  + \psi_{ij}(\tilde{\bm k}) \,,\label{eq:subensemble_spectrum_characteristic_evo}
\end{align}
where we dropped the arguments of $\tilde{\bm k}(\bm q, t)$ to lighten notation. In the last step, we inserted~\eqref{eq:subensemble_spectrum_evo} and used incompressibility $k_p \phi^{\bm v}_{pq}(\bm k, t) = k_q \phi^{\bm v}_{pq}(\bm k, t) = 0$. By following the characteristics, we have thus removed the gradient term. The spectrum along each characteristic evolves independently from the other ones. 

In the same way, the method of characteristics can be used to remove the gradient term in~\eqref{eq:scalar_spec_tensor_evo},
\begin{align} \label{eq:scalar_spec_tensor_characteristic_evo}
  \partial_t \tilde{\phi^{\bm v}}(\bm q, t)
  &= - 2\nu \tilde{k}^2 \tilde{\phi^{\bm v}}(\bm q, t)
   + \frac{Q(\tilde{k})}{4\pi\tilde{k}^2} \,,
\end{align}
showing that most of the model dynamics is captured by the dynamically evolving wavevectors, with only the viscous and forcing contributions acting additionally along each characteristic.

\subsubsection{Logarithmically scaled, dynamic grid}
Since the dynamics of the system is such that energy is transported from the large scales to the small scales, we need to include large wavenumbers, particularly when considering small values of the viscosity $\nu$. In order to achieve this in our numerical implementation, we construct a grid of wavevectors that is spaced logarithmically in radial direction. Let $\bm g \in (\Delta g \mathbb{Z})^3$ be a point on a regular grid with grid spacing $\Delta g$, then we compute the grid points of the wavevectors as
\begin{align}
  \bm q(\bm g) = q_0 \exp(g) \hat{\bm g}\,,
\end{align}
where $g = |\bm g|$, $\hat{\bm g} = \bm g/g$, and $q_0$ is some scaling factor. Combined with the dynamic evolution of the grid, we get dynamically evolving wavevectors
\begin{align} \label{eq:grid_map}
  \tilde{k}_i(\bm g, t) = W_{ji}(t) q_j(\bm g) = q_0 \exp(g) W_{ji}(t) \hat{g}_j\,.
\end{align}
Since each characteristic evolves independently, we can simulate the evolution of the field along each of these dynamic wavevectors using~\eqref{eq:subensemble_spectrum_characteristic_evo} (or~\eqref{eq:scalar_spec_tensor_characteristic_evo}).

Another issue that arises due to the dynamic evolution of the grid comes from the fact that, over time, the deformation tensor $\mathrm{W}$ becomes more and more ill-conditioned (mapping a sphere to a strongly elongated ellipsoid)~\cite[p.~919]{FalkovichRMP2001}. Since we want to explore the statistically stationary state of the system that emerges only after a certain transient time, we need to reset the deformation tensor to $W_{ij} = \delta_{ij}$ in regular time intervals. This changes the position of the discrete wavevectors $\tilde{\bm k}$ and thus requires remapping, i.e., interpolating the $\phi^{\bm v}_{ij}(\bm k, t)$ field at these new positions. Thanks to the smooth map~\eqref{eq:grid_map} (apart from the point $\bm g = 0$), we can understand the field $\tilde{\phi}^{\bm v}_{ij}(\bm q(\bm g), t)$ as defined on a regular grid $\{\bm g \in (\Delta g \mathbb{Z})^3\}$. On this square grid, we use spline interpolation with continuous 2nd-order derivatives~\cite{lalescu2010} to remap the spectrum tensor to the new grid points. The remapping is performed every $n_\text{remap}$ iterations of the simulations.

\subsubsection{Time stepping}
For the description of the time stepping, we here only discuss the case of the scalar equation~\eqref{eq:scalar_spec_tensor_characteristic_evo}. Time stepping for the tensorial field with general $\gamma$ is described in Appendix~\ref{sec:numerics_general_gamma}.
We define an integrating factor,
\begin{align}
  A(\bm q, t) = 2\nu \int_0^{t} \dif s \,\tilde{k}(\bm q, s)^2\,,
\end{align}
which we use to define the compensated spectrum
\begin{align} \label{eq:def_compensated_spectrum}
  Z(\bm q, t) &= e^{A(\bm q, t)} \tilde{\phi}^{\bm v}(\bm q, t)\,.
\end{align}
Then, the three equations that need to be solved simultaneously are
\begin{align}
  \dod{}{t} W_{ij}(t) &= - W_{il}(t) B_{lj}(t) \label{eq:numsolver_eq_1}\\
  \dpd{}{t} A(\bm q, t) &= 2\nu \tilde{k}(\bm q, t)^2  \label{eq:numsolver_eq_2}\\
  \dpd{}{t} Z(\bm q, t) &= e^{A(\bm q, t)} \frac{Q(\tilde{k}(\bm q, t))}{4\pi\tilde{k}(\bm q, t)^2} \,.  \label{eq:numsolver_eq_3}
\end{align}
In the last equation, the viscous term disappeared thanks to the integrating factor. The deformation tensor $\mathrm{W}$ evolves independently of the spectrum dynamics, whereas the integrating factor $A$ and the compensated spectrum $Z$ depend on $\mathrm{W}$ through the dynamic wavevectors. They need to be simulated separately for each wavevector $\bm q$.

We use a Stratonovich-type, explicit method~(see References~\cite[Euler-Heun]{RackauckasJORS2017} and \cite{roberts2012}). For a time step $\Delta t$ we compute an intermediate step
\begin{align}
  W_{ij}'(t) &= W_{ij}(t) - W_{il}(t) \overline{B_{lj}}(t) \sqrt{\Delta t} \label{eq:deformationtensor_predictor_step}\\
  A'(\bm q, t) &= A(\bm q, t) + 2\nu \tilde{k}(\bm q, t)^2 \Delta t \\
  Z'(\bm q, t) &= Z(\bm q, t) + 
  e^{A(\bm q, t)} \frac{Q(\tilde{k}(\bm q, t))}{4\pi\tilde{k}(\bm q, t)^2}\Delta t
\end{align}
and then the final step
\begin{align}
  W_{ij}(t + \Delta t) &= W_{ij}(t) - \frac12 \left[W_{il}(t) + W_{il}'(t)\right] \overline{B_{lj}}(t) \sqrt{\Delta t}  \label{eq:deformationtensor_corrector_step} \\
  A(\bm q, t+\Delta t) &= A(\bm q, t) + \frac12 \left[2\nu \tilde{k}(\bm q, t)^2 + 2\nu {\tilde{k}'(\bm q, t)}^2 \right] \Delta t \\
  Z(\bm q, t+\Delta t) &= Z(\bm q, t) + \frac12 \left[e^{A(\bm q, t)}\frac{Q(\tilde{k}(\bm q, t))}{4\pi\tilde{k}(\bm q, t)^2} + e^{A'(\bm q, t)} \frac{Q(\tilde{k}'(\bm q, t))}{4\pi\tilde{k}'(\bm q, t)^2}\right] \Delta t \,,
\end{align}
where $\tilde{\bm k}'(\bm q, t) = \mathrm{W}'(t)^T \bm q$ and $\overline{\mathrm{B}}(t)$ is a zero-mean Gaussian matrix with correlations (compare eq.~\eqref{eq:b_covariance})
\begin{align}
  \left\langle \overline{B_{ik}}(t) \overline{B_{jl}}(t) \right\rangle &= \beta \left( \delta_{ij}\delta_{kl} - \frac{1}{4} \delta_{ik}\delta_{jl} - \frac{1}{4} \delta_{il}\delta_{jk} \right) \,,
\end{align}
generated independently for each time step. This can be rewritten as a scheme for $\tilde{\phi}^{\bm v}(\bm q, t)$ as
\begin{align}
  \tilde{\phi}^{\bm v}(\bm q, t + \Delta t) &= e^{-A(\bm q, t + \Delta t)} Z(\bm q, t + \Delta t) \\
  &= e^{-\Delta A(\bm q, t)} \tilde{\phi}^{\bm v}(\bm q, t)
  + \frac12 e^{-\Delta A(\bm q, t)} \left[\frac{Q(\tilde{k}(\bm q, t))}{4\pi\tilde{k}(\bm q, t)^2} + e^{\Delta A'(\bm q, t)} \frac{Q(\tilde{k}'(\bm q, t))}{4\pi\tilde{k}'(\bm q, t)^2}\right] \Delta t \label{eq:field_timestep}
\end{align}
where we defined
\begin{align}
  \Delta A'(\bm q, t) &\equiv A'(\bm q, t) - A(\bm q, t) = 2\nu \tilde{k}(\bm q, t)^2 \Delta t \\
  \Delta A(\bm q, t) &\equiv A(\bm q, t+\Delta t) - A(\bm q, t) = \frac12 \left[2\nu \tilde{k}(\bm q, t)^2 + 2\nu {\tilde{k}'(\bm q, t)}^2 \right] \Delta t\,.
\end{align}
Since only the $A$-increments are needed, we do not need to save the values of $A$. Furthermore, it turns out that for solving these simplified scalar equations, we can even content ourselves with computing the deformation tensor steps~\eqref{eq:deformationtensor_predictor_step} and \eqref{eq:deformationtensor_corrector_step}, and then the field can be computed in the single step~\eqref{eq:field_timestep}. This is because the right-hand side of~\eqref{eq:numsolver_eq_3} does not depend on $Z$. The numerics for the general case $\gamma \neq 0$ is described in Appendix~\ref{sec:numerics_general_gamma} along with an example simulation presented in Appendix~\ref{sec:numerical_results_small_gamma}.

\subsection{Numerical results}

\begin{table}[tb]
  \caption{Parameters of the ensemble simulations.
    \label{tab:sim_params}}
  \begin{ruledtabular}
  \begin{tabular}{llllllllll}
  grid size & \# members & $\beta$ & $\nu$ & $\gamma$ & $Q(k)$ & $q_0$ & $\Delta g$ & $\Delta t$ & $n_\mathrm{remap}$ \\
  $256^3$ & 40 & 16 & $10^{-3}$ & 0 & $2\pi k^4 \exp(-(k/2)^2)$ & 0.1 & 0.075 & $5\times 10^{-4}$ & 100
  \end{tabular}
  \end{ruledtabular}
\end{table}
\begin{figure}[tb]
  \includegraphics{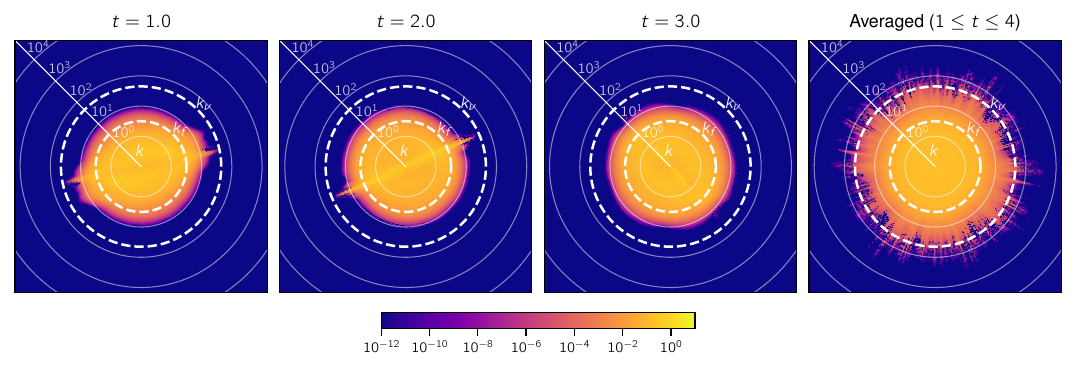}
  \caption{Slices through the trace of the spectrum tensor $\phi_{ii}^{\bm v}(\bm k, t)$ in the ($k_y=0$)-plane for one of the ensemble members in the case $\gamma=0$. The first three panels are snapshots at three different times $t$. The last panel corresponds to an average over all snapshots in the interval $1 \leq t \leq 4$. The magnitude of $\bm k$ is spaced logarithmically. A forcing wavenumber $k_f = \mathrm{argmax}[k Q(k)]$ and a viscous wavenumber $k_\nu = \lambda^{-1}$ are indicated by dashed lines.}
  \label{fig:excursions}
\end{figure}

We implemented a solver for the SPDE~\eqref{eq:subensemble_spectrum_evo} as described in the previous section in Julia~\cite{BezansonSR2017}. In the following, we present numerical results based on an ensemble of simulations with parameters listed in table~\ref{tab:sim_params}. The initial condition is set to zero, i.e., $\phi^{\bm v}_{ij}(\bm k, 0) \equiv 0$. To illustrate the dynamics of these simulations, figure~\ref{fig:excursions} shows slices of the trace of the spectrum $\phi^{\bm v}_{ii}(\bm k, t)$ in the ($k_y = 0$)-plane at different times $t$ of one of the simulations. The magnitude of the $\bm k$-vectors is spaced logarithmically, corresponding to how $\bm k$-space is represented also numerically. Due to the forcing, energy is injected at low $k$-values, corresponding to the center of the plots. Over time, the stochastic stretching transports this energy by distorting $\bm k$-space according to~\eqref{eq:characteristic_evo}, leading to intermittent excursions of energy into regions with larger $k$, where it is removed again by viscosity. While being anisotropic instantaneously, the time-averaged panel shows that the excursions are statistically isotropic. Next, we characterize the statistical properties of the system.

\begin{figure}[tb]
  \includegraphics{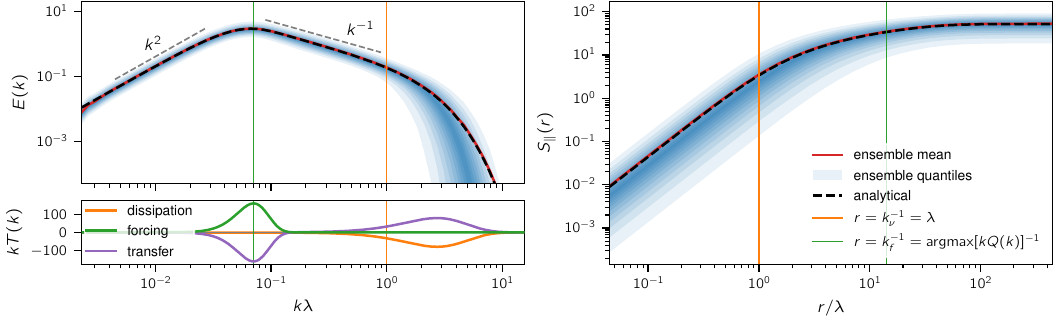}
  \caption{Mean energy across scales ($\gamma=0$).
  Top left: Mean energy spectrum of the ensemble simulations (red, solid) compared to the analytical solution~\eqref{eq:complete_solution_spectral_energy_budget_gamma=0} (black, dashed). The blue, shaded area indicates the distribution of the energy spectrum across snapshots of the ensemble of simulations in the stationary state, each level set corresponding to quantiles incremented by $5\%$.
  Bottom left: Dissipation, forcing, and transfer term (represented by the label $T(k)$) in the spectral energy budget equation~\eqref{eq:spectral_energy_budget_gamma=0}, computed for the analytical solution~\eqref{eq:complete_solution_spectral_energy_budget_gamma=0}. They are multiplied by $k$ so that the area below the curve corresponds to a rate of energy injection/dissipation in the semi-logarithmic plot.
  Right: Mean second-order longitudinal structure function. Colors and line styles are consistent with the top left panel.}
  \label{fig:spec1d}
\end{figure}

As a starting point, we consider the second-order statistics. In section~\ref{sec:spectral_energy_budget}, we computed the stationary solution~\eqref{eq:complete_solution_spectral_energy_budget_gamma=0} of the mean spectral energy budget equation~\eqref{eq:spectral_energy_budget_gamma=0}. In figure~\ref{fig:spec1d} (top left), we compare this solution (black, dashed) to the average energy spectrum from our ensemble simulations (red, solid), which coincide very well. They feature $k^2$ scaling at the large scales and $k^{-1}$ scaling in the inertial range. In figure~\ref{fig:spec1d} (bottom left), we compute the forcing term, the viscous term, and the transfer term of~\eqref{eq:spectral_energy_budget_gamma=0} to illustrate how energy flows through the system. The forcing term injects energy at small wavenumbers. On average, the transfer term removes this energy from the small wavenumbers and transports it to large wavenumbers, where it is dissipated by the viscous term. In order to represent rates of energy flow visually by the areas under the curves, the terms in this semi-logarithmic plot are multiplied by $k$. An analogous representation of the energy across scales can be given by the longitudinal second-order structure function. In each sub-ensemble, it can be computed as (compare~\cite[pp.~224--227]{pope2000}):
\begin{align}
  S^{\bm v}_\parallel(r, t) &= \left\langle (u_1(\bm x + r\bm e_1, t) - u_1(\bm x, t))^2 \right\rangle_{\bm F}
  \label{eq:sub-ensemble_strucfunc}\\
  &= 2\int \dif \bm k \left(1-\cos(k_1 r)\right) \phi^{\bm v}_{11}(\bm k, t)\,.
  \label{eq:sub-ensemble_strucfunc_from_spectrum_tensor}
\end{align}
Note that $S^{\bm v}_\parallel(r, t)$ is a fluctuating quantity across the sub-ensembles.
For the full ensemble, isotropy allows us to rewrite the structure function as
\begin{align}
  S_\parallel(r, t) &= \left\langle S^{\bm v}_\parallel(r, t) \right\rangle_{\bm v}
  = \int_0^\infty \dif k\, E(k, t) \left(\frac43 - \frac{4}{(kr)^3}\sin(kr) + \frac{4}{(kr)^2}\cos(kr)\right) \,.
\end{align}
In figure~\ref{fig:spec1d} (right), the mean structure function computed from the analytical solution (black, dashed) is compared to the average from the ensemble simulations (red, solid), coinciding perfectly. 

Each snapshot from the ensemble simulations of the SPDE~\eqref{eq:subensemble_spectrum_evo} represents a realization of the sub-ensemble spectrum tensor $\phi^{\bm v}_{ij}(\bm k, t)$, where we have averaged over the stochastic forcing (for deriving~\eqref{eq:subensemble_spectrum_evo}) but not over the stochastic gradients $\mathrm{B}$. As we showed in section~\ref{sec:stat_field_theory}, each sub-ensemble conditional on the time series of $\mathrm{B}(t)$ remains Gaussian, and thus its statistics at each point in time $t$ are comprehensively characterized by the sub-ensemble spectrum tensor $\phi^{\bm v}_{ij}(\bm k, t)$. Non-Gaussianity in our model is generated by stochasticity of $\mathrm{B}$ and thus variations of the spectrum tensor $\phi^{\bm v}_{ij}(\bm k, t)$ across the different sub-ensembles. Therefore, characterizing the statistics of the spectrum tensor as a solution to the SPDE~\eqref{eq:subensemble_spectrum_evo} directly translates into characterizing the non-Gaussianity of the model. These variations of the spectrum tensor are shown as shades of blue in figure~\ref{fig:spec1d}. For figure~\ref{fig:spec1d} (top left), for example, we determine them by computing the energy spectrum at each snapshot of each realization of the system in the stationary state. This provides a distribution of energy spectra, whose quantiles are shown as shades of blue. The quantiles reveal that fluctuations (and thus non-Gaussianity) grow toward the smaller scales. We also observe that the mean values (red lines) of both the energy spectrum and the structure function visibly deviate from the median values (darkest shade of blue) at small scales, pointing to a skewed distribution of these quantities across the ensemble. Note that in the two other physically relevant cases $\gamma=\pm 1$, the field grows without bounds, as discussed in Appendix~\ref{sec:spectral_budget_general_gamma}. As a result, the model does not converge to a stationary state. Instead, in Appendix~\ref{sec:numerical_results_small_gamma}, we show analogous numerical results for a simulation with $\gamma= 0.5$ (which is still convergent) to illustrate that our analysis and numerics generalize to $\gamma\neq 0$ as long as there is a stationary state.

\begin{figure}[tb]
  \includegraphics{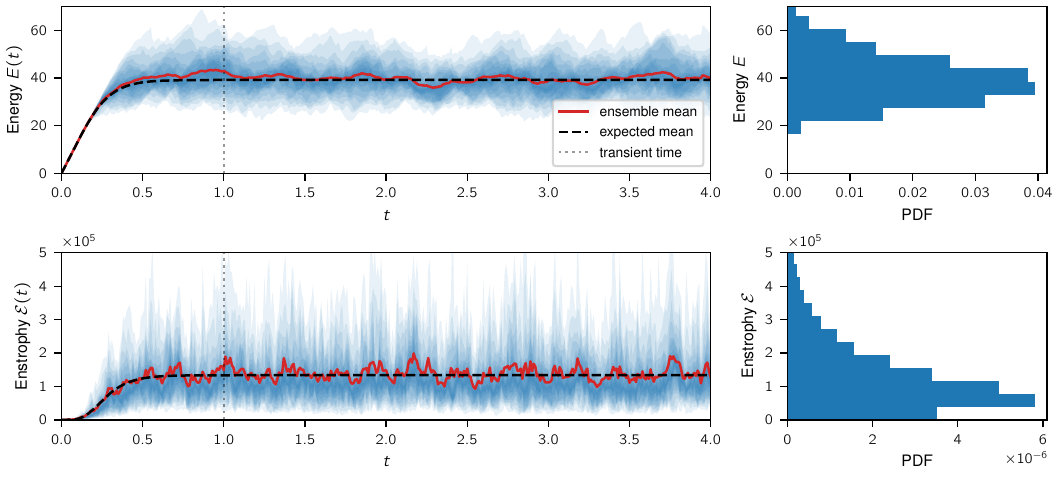}
  \caption{Energy and enstrophy of the model ($\gamma=0$). Left panels: Energy and enstrophy in the ensemble simulations over time, starting at zero and then saturating at a statistically stationary state. The mean over the 40 ensemble simulations (solid, red) fluctuates around the numerical solution of the spectral energy budget equation~\eqref{eq:spectral_energy_budget_gamma=0} (black, dashed). The blue, shaded area indicates the distribution across the ensemble of simulations, each level set corresponding to quantiles incremented by $5\%$. A transient $T_\mathrm{tr}=1$ (grey, dotted) is chosen manually to demarcate the beginning of the stationary state. Right panels: Distribution of energy and enstrophy in the stationary state.}
  \label{fig:energy_enstrophy}
\end{figure}

Here, we focus on the case $\gamma=0$. Figure~\ref{fig:energy_enstrophy} (left panels) shows the total energy and enstrophy in our simulations over time, computed as
\begin{align}
  E(t) &= \frac{1}{2} \int \dif\bm k\, \phi_{ii}(\bm k, t)
\end{align}
and
\begin{align}
  \mathcal{E}(t) &= \int \dif\bm k\, k^2 \phi_{ii}(\bm k, t)\,.
\end{align}
The average over the ensemble simulations (red, solid) is compared to the value computed from a numerical solution of the spectral energy budget equation~\eqref{eq:spectral_energy_budget_gamma=0} (black, dashed). The variations across the ensemble simulations are shown in shades of blue. While all simulations are launched with zero initial condition, it appears that they converge to a statistically stationary state. Based on this, we choose a transient time $T_\mathrm{tr}=1$. To characterize the stationary state, we consider all snapshots at $t\geq T_\mathrm{tr}$ as samples, which were also the ones used for figure~\ref{fig:spec1d}. The right panels of figure~\ref{fig:energy_enstrophy} show the distribution of energy and enstrophy across realizations of $\phi^{\bm v}_{ij}(\bm k, t)$ in the stationary state. They confirm that the energy (a more large-scale quantity) displays a fairly narrow distribution while the distribution of enstrophy (a more small-scale quantity) is very broad and skewed.

\begin{figure}[tb]
  \includegraphics{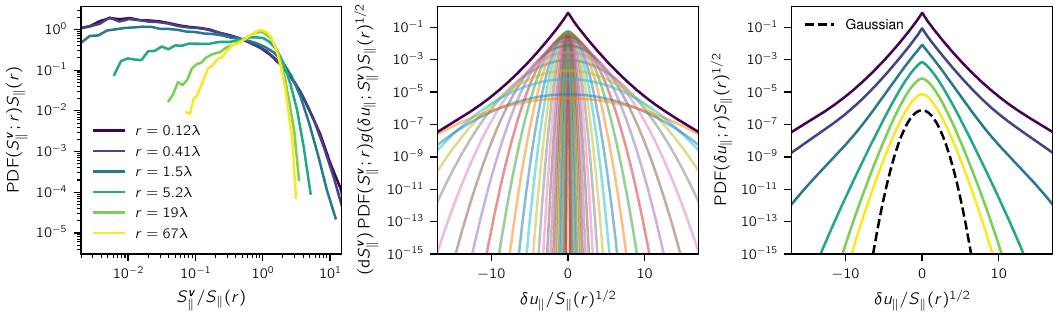}
  \caption{How to construct longitudinal increment statistics ($\gamma=0$). Left: Distribution~\eqref{eq:subens_struc_func_pdf} of the second-order structure function $S^{\bm v}_\parallel(r,t)$ for different $r$ across realizations of $\phi^{\bm v}_{ij}(\bm k, t)$, computed by histograms of~\eqref{eq:sub-ensemble_strucfunc_from_spectrum_tensor}. Center: Gaussian PDFs (colored lines) used to construct the increment PDF of the full ensemble at $r=0.12\lambda$ by the superposition~\eqref{eq:gauss_incr_pdf_superpos}. Each Gaussian PDF corresponds to a value of ${S^{\bm v}_\parallel}$ and is weighted like the integrand in~\eqref{eq:gauss_incr_pdf_superpos}. The resulting integrated PDF is also shown (violet). Right: Standardized longitudinal increment PDFs of the full model~\eqref{eq:passive_vector_model} for different values of $r$, compared to a Gaussian PDF (black, dashed), vertically shifted for clarity.}
  \label{fig:increments}
\end{figure}

Finally, we can use the ensemble simulations to compute detailed statistics of the full model~\eqref{eq:passive_vector_model}. For example, we know that longitudinal velocity increments $\delta u_\parallel(r, t) = u_1(r\bm e_1, t) - u_1(\bm 0, t)$ have (zero-mean) Gaussian statistics in each sub-ensemble, since they are linear functionals of the field $\bm u(\bm x, t)$. So the PDF of increments in sub-ensemble $\bm v$ reads
\begin{align}
  \mathrm{PDF}^{\bm v}(\delta u_\parallel; r, t) &\equiv \left\langle \delta(\delta u_\parallel(r, t) - \delta u_\parallel) \right\rangle_{\bm F} \\
  &= g(\delta u_\parallel; S^{\bm v}_\parallel(r, t)) \,, \label{eq:gaussian_subens_incr_pdf}
\end{align}
where $\delta u_\parallel$ is the sample space variable of the increments (as opposed to the realization $\delta u_\parallel(r, t)$) and
\begin{align}
  g(x; \sigma^2) &= \frac{1}{\sqrt{2\pi \sigma^2}} \exp\left(-\frac{x^2}{2 \sigma^2}\right)
\end{align}
is a Gaussian PDF. Here, the variance given by the second-order structure function $S^{\bm v}_\parallel(r, t)$~\eqref{eq:sub-ensemble_strucfunc} fully characterizes the increment distribution in each sub-ensemble. To compute the increment PDF of the full ensemble, an additional average over the realizations of $\bm v$ is required. This can be written as an integral over the values of $S^{\bm v}_\parallel(r, t)$,
\begin{align}
  \mathrm{PDF}(\delta u_\parallel; r) &\equiv \left\langle \delta(\delta u_\parallel(r, t) - \delta u_\parallel) \right\rangle \\
  &= \left\langle \mathrm{PDF}^{\bm v}(\delta u_\parallel; r, t) \right\rangle_{\bm v} \\
  &= \int \dif {S^{\bm v}_\parallel} \left\langle \delta(S^{\bm v}_\parallel(r,t) - {S^{\bm v}_\parallel}) \right\rangle_{\bm v} g(\delta u_\parallel; {S^{\bm v}_\parallel}) \\
  &= \int \dif {S^{\bm v}_\parallel}\,\mathrm{PDF}({S^{\bm v}_\parallel}; r) g(\delta u_\parallel; {S^{\bm v}_\parallel}) \label{eq:gauss_incr_pdf_superpos}
\end{align}
where we introduced the sample space variable of the sub-ensemble structure function ${S^{\bm v}_\parallel}$ along with its PDF
\begin{align} \label{eq:subens_struc_func_pdf}
  \mathrm{PDF}({S^{\bm v}_\parallel}; r) \equiv \left\langle \delta(S^{\bm v}_\parallel(r,t) - {S^{\bm v}_\parallel}) \right\rangle_{\bm v}\,.
\end{align}
We also dropped the time dependence for the full ensemble by assuming that we are in the statistically stationary state. The superposition~\eqref{eq:gauss_incr_pdf_superpos} is visualized in figure~\ref{fig:increments}. The left panel shows the PDF~\eqref{eq:subens_struc_func_pdf} of values of the sub-ensemble structure function for different distances $r$. The width of the distribution increases at smaller distances, corresponding to an increase in intermittency toward the smaller scales. The middle panel illustrates the superposition of Gaussian PDFs given by~\eqref{eq:gauss_incr_pdf_superpos}. Each of the colored lines represents a Gaussian PDF, weighted by the probability of the bin for computing the integral~\eqref{eq:gauss_incr_pdf_superpos} at $r=0.12\lambda$. The resulting sum/integral of these Gaussians is also shown (violet). Finally, the right panel shows the full increment PDFs resulting from the superposition~\eqref{eq:gauss_incr_pdf_superpos} at different $r$, including the one constructed in the middle panel. They display a transition from heavy-tailed statistics at small scales to close-to-Gaussian at large scales as typically observed in turbulence, albeit without any skewness.


\section{Conclusions}

We investigate a class of passive vector models~\eqref{eq:passive_vector_model} with a temporally delta-correlated and spatially linear incompressible advecting field. This is similar to the Batchelor regime of passive scalars, in which the scalar develops variations well below the scale at which the velocity becomes smooth. Thanks to the linearity, the model proves to be very tractable analytically. We derive the evolution equation for the mean energy spectrum as well as its exact stationary solution (for $\gamma$ equal to and around $0$). Consistent with the literature on the kinematic dynamo~\cite{KazantsevSPJ1968,VergassolaPRE1996,AdzhemyanE2001,VincenziJoSP2002,ArponenJSP2007} and the linearized Navier-Stokes equation~\cite{AdzhemyanE2001,ArponenPRE2010}, the model becomes unstable for values of $\gamma$ away from $0$.

We consider the Hopf functional calculus for the model and find that its statistics are described by a Hopf equation with an advection term featuring a second-order functional derivative, reminiscent of the inertial term for Navier-Stokes turbulence. As in Navier-Stokes, this term generates intermittency with an increasing degree of non-Gaussianity toward the smaller scales. In our model, however, we show that the Hopf equation can be solved exactly by a superposition of statistically homogeneous but anisotropic Gaussian characteristic functionals, which represent statistics conditional on realizations of the advecting field. 

Our numerical simulations make use of this Gaussian decomposition. They illustrate how sudden spikes of transfer of energy toward smaller scales accumulate over time (see figure~\ref{fig:excursions}) and finally form broad distributions of energy at the small scales, which---due to the Gaussian decomposition---generate heavy-tailed statistics at the small scales (see figure~\ref{fig:increments}). Thus, our analytically tractable model allows to dissect and illustrate a mechanism for generating small-scale intermittency, reminiscent of the complex processes involved in generating intermittency in Navier-Stokes turbulence.

\subsection*{Acknowledgments}
We would like to thank Theodore D.~Drivas for pointing out to us the analogy to the Batchelor limit of passive scalar advection. We thank Maurizio Carbone and Gabriel Brito Apolin\'ario for interesting and helpful discussions in the course of this project.
We thank the anonymous reviewers for the exceptionally detailed feedback, which contributed significantly to improving the manuscript.
This project has received funding from the European Research Council (ERC) under the European Union's Horizon 2020 research and innovation programme (Grant agreement No.\ 101001081).
The authors gratefully acknowledge the scientific support and HPC resources provided by the Erlangen National High Performance Computing Center (NHR@FAU) of the Friedrich-Alexander-Universit\"at Erlangen-N\"urnberg (FAU) under the NHR project \texttt{b159cb}. NHR funding is provided by federal and Bavarian state authorities. NHR@FAU hardware is partially funded by the German Research Foundation (DFG) -- 440719683. Mathematica and ChatGPT were used to assist in the analysis of the spectral energy budget equations.

\appendix

\section{Derivation of the spectral energy budget equation}
\label{sec:energy_budget_derivation}

The spectral energy tensor $\phi_{ij}(\bm k, t)$ for a statistically homogeneous field is related to the Fourier covariance through
\begin{equation}
  \label{eq:passive_vector_fourier_covariance}
  \langle \hat{u}_i(\bm k,t) \hat{u}^*_j(\bm k',t) \rangle = \phi_{ij}(\bm k,t) \delta(\bm k - \bm k') \, , 
\end{equation}  
  where the star denotes complex conjugation (see \eqref{eq:spectral_energy_tensor_general} and \eqref{eq:spectral_energy_tensor}). To derive an evolution equation for the energy spectrum function $E(k,t)$, we therefore consider
\begin{align}
  \partial_t \langle \hat{u}_i(\bm k,t)\hat{u}^*_i(\bm k,t) \rangle = &-2 \nu k^2 \langle \hat{u}_i(\bm k,t)\hat{u}^*_i(\bm k,t) \rangle + \langle \hat{u}_i(\bm k,t) \hat{F}^*_i(\bm k,t) \rangle + \langle \hat{F}_i(\bm k,t) \hat{u}^*_i(\bm k,t) \rangle \nonumber \\
  &+ \langle k_l B_{lm}(t) \hat{u}^*_j(\bm k,t) \partial_m \hat{u}_j(\bm k,t) \rangle + \langle \gamma \hat{u}^*_j(\bm k,t) B_{jm}(t) \hat{u}_m(\bm k,t) \rangle \nonumber  \\
  &+ \langle k_l B_{lm}(t) \hat{u}_j(\bm k,t) \partial_m \hat{u}^*_j(\bm k,t) \rangle + \langle \gamma \hat{u}_j(\bm k,t) B_{jm}(t) \hat{u}^*_m(\bm k,t) \rangle \,. \label{eq:covariance_evolution}
\end{align}
To evaluate the averages involving $\mathrm{B}$ and $\hat{\bm F}$, we make use of Gaussian integration by parts \cite{novikov1965,FurutsuJRNBSD1963,DonskerAFS1964}.
For the term related to the forcing, we need to evaluate
\begin{equation}
  \langle \hat{u}_i(\bm k,t) \hat{F}^*_i(\bm k,t) \rangle 
  = \int \dif\bm k'\int \dif s \, \langle \hat{F}^*_i(\bm k,t) \hat{F}_j(\bm k',s) \rangle \left\langle \frac{\delta \hat{u}_i(\bm k,t)}{\delta \hat{F}_j(\bm k',s)} \right\rangle \,.
\end{equation}
As specified in~\eqref{eq:forcing_covariance} and \eqref{eq:forcing_spectrum}, the forcing Fourier covariance takes the isotropic form
\begin{equation}
  \langle \hat{F}_i(\bm k,t) \hat{F}^*_j(\bm k',t') \rangle = \psi_{ij}(\bm k) \delta(\bm k - \bm k') \delta(t-t') = \frac{Q(k)}{4\pi k^2} P_{ij}(\bm k) \delta(\bm k - \bm k') \delta(t-t') \, ,
\end{equation}
which leads to
\begin{equation}
  \langle \hat{u}_i(\bm k,t) \hat{F}^*_i(\bm k,t) \rangle =  \frac{Q(k)}{4\pi k^2} P_{ij}(\bm k) \left\langle \frac{\delta \hat{u}_i(\bm k,t)}{\delta \hat{F}_j(\bm k,t)} \right\rangle \, .
\end{equation}
To compute the mean response function $\langle \delta \hat{u}_i(\bm k,t) / \delta \hat{F}_j(\bm k,t) \rangle$, we first write down the formal solution of \eqref{eq:passive_vector_model_fourier}:
\begin{align}
  \hat{u}_i(\bm k,t) &= \hat{u}_i(\bm k,0) \nonumber \\
  &+ \int_0^t \dif s \, \left( \left[- \mathrm i k_l V_l(s) - \nu k^2 \right] \hat{u}_i(\bm k,s) + P_{ij}(\bm k) \left[ k_l B_{lm}(s) \partial_m \hat{u}_j(\bm k,s)+ \gamma B_{jm}(s) \hat{u}_m(\bm k,s) \right] + \hat{F}_i(\bm k,s) \right) \,.
\end{align}
Based on that, we can compute the mean response function for the forcing,
\begin{equation}
  \left\langle \frac{\delta \hat{u}_i(\bm k,t)}{\delta \hat{F}_j(\bm k',t)} \right\rangle = \frac{1}{2} \delta_{ij} \delta(\bm k - \bm k') \,,
\end{equation}
where the factor $1/2$ comes from integrating a delta function at the endpoint of the integration interval.
We thus obtain
\begin{equation}
  \langle \hat{u}_i(\bm k,t) \hat{F}^*_i(\bm k,t) \rangle = \langle \hat{u}^*_i(\bm k,t) \hat{F}_i(\bm k,t) \rangle = \frac{Q(k)}{4\pi k^2} \delta(\bm 0) \,.
  \label{eq:gipb_force}
\end{equation}
Note that the divergent factor $\delta(\bm 0)$ will ultimately cancel out since it is also contained in all other terms. 
For the remaining terms, we need to compute averages of the form
\begin{align}
  \langle \hat{u}^*_j(\bm k,t) B_{lm}(t) \hat{u}_k(\bm k,t) \rangle &= \int \dif s \left\langle B_{lm}(t)B_{np}(s) \right\rangle \left\langle \frac{\delta [ \hat{u}^*_j(\bm k,t) \hat{u}_k(\bm k,t) ]}{\delta B_{np}(s)} \right\rangle \\
  &=  \beta \left( \delta_{ln}\delta_{mp} - \frac{1}{4} \delta_{lm}\delta_{np} - \frac{1}{4} \delta_{lp}\delta_{mn} \right) \left\langle \frac{\delta [ \hat{u}^*_j(\bm k,t) \hat{u}_k(\bm k,t) ]}{\delta B_{np}(t)} \right\rangle \,,
\end{align}
where we have inserted \eqref{eq:b_covariance}.
For the mean response function we obtain
\begin{align}
  \left\langle \frac{\delta [ \hat{u}^*_j(\bm k,t) \hat{u}_k(\bm k,t) ]}{\delta B_{np}(t)} \right\rangle &= \frac{1}{2} \left\langle P_{jq} \hat{u}_k k_n \partial_p \hat{u}^*_q + P_{kq} \hat{u}^*_j k_n \partial_p \hat{u}_q + \gamma P_{jn} \hat{u}_k \hat{u}^*_p + \gamma P_{kn} \hat{u}_p \hat{u}^*_j \right\rangle\,,
\end{align}
leaving out the arguments to ease notation. Contracting the Kronecker deltas results in
\begin{align}
  \langle \hat{u}^*_j(\bm k,t) B_{lm}(t) \hat{u}_k(\bm k,t) \rangle &= \frac{1}{2}\beta \langle P_{jq} \hat{u}_k k_l \partial_m \hat{u}^*_q + P_{kq} \hat{u}^*_j k_l \partial_m \hat{u}_q + \gamma P_{jl} \hat{u}_k \hat{u}^*_m + \gamma P_{kl} \hat{u}_m \hat{u}^*_j \rangle \nonumber \\
  &-\frac{1}{8}\beta \delta_{lm} \langle P_{jq} \hat{u}_k k_n \partial_n \hat{u}^*_q + P_{kq} \hat{u}^*_j k_n \partial_n \hat{u}_q + \gamma P_{jn} \hat{u}_k \hat{u}^*_n + \gamma P_{kn} \hat{u}_n \hat{u}^*_j \rangle \nonumber \\
  &-\frac{1}{8}\beta \langle P_{jq} \hat{u}_k k_m \partial_l \hat{u}^*_q + P_{kq} \hat{u}^*_j k_m \partial_l \hat{u}_q + \gamma P_{jm} \hat{u}_k \hat{u}^*_l + \gamma P_{km} \hat{u}_l \hat{u}^*_j \rangle \, .
\end{align}

With this, we can evaluate the averages in \eqref{eq:covariance_evolution}. We get
\begin{align}
&\quad\gamma \langle \hat{u}^*_j(\bm k,t) B_{jm}(t) \hat{u}_m(\bm k,t) + \hat{u}_j(\bm k,t) B_{jm}(t) \hat{u}^*_m(\bm k,t) \rangle \\
&= \frac{7}{4} \beta \gamma^2 \langle \hat{u}_j \hat{u}^*_j \rangle- \frac{1}{4} \beta \gamma k_n \partial_n \langle \hat{u}_j \hat{u}^*_j \rangle\\
&= \left[ \frac{7}{4} \beta \gamma^2 \frac{E(k,t)}{2\pi k^2} - \frac{1}{4} \beta \gamma k \frac{\partial}{\partial k} \frac{E(k,t)}{2\pi k^2} \right]\delta(\bm 0) \\
&= \frac{\beta}{2 \pi k^2} \left[ \frac{7}{4} \gamma^2 E(k,t) - \frac{1}{4} \gamma k \frac{\partial E(k,t)}{\partial k} + \frac{1}{2} \gamma E(k,t) \right] \delta(\bm 0)
\label{eq:gipb_gamma_term}
\end{align}
as well as
\begin{align}
  &\quad \langle k_l B_{lm}(t) \hat{u}^*_j(\bm k,t) \partial_m \hat{u}_j(\bm k,t) + k_l B_{lm}(t) \hat{u}_j(\bm k,t) \partial_m \hat{u}^*_j(\bm k,t) \rangle \\
  &= k_l \partial_m \langle B_{lm}(t) \hat{u}_j(\bm k,t) \hat{u}^*_j(\bm k,t) \rangle \\
  &= -\frac{1}{4} \beta \gamma k_l \partial_l \langle \hat{u}_j \hat{u}^*_j \rangle + \frac{1}{2} \beta k^2 \partial_l\partial_l \langle \hat{u}_j \hat{u}^*_j \rangle - \frac{1}{4} \beta k_l k_n \partial_l \partial_n \langle \hat{u}_j \hat{u}^*_j \rangle + \frac{3}{8} \beta \gamma k_l \partial_m \langle \hat{u}_l \hat{u}^*_m + \hat{u}^*_l \hat{u}_m \rangle \\
  &= \frac{\beta}{4} \left[ (4-\gamma) k \dpd{}{k} \frac{E(k, t)}{2\pi k^2} - 3 \gamma \frac{E(k, t)}{2\pi k^2} + k^2 \dpd[2]{}{k} \frac{E(k, t)}{2\pi k^2} \right]\delta(\bm 0)\\
  &= \frac{\beta}{2 \pi k^2} \left[ - \frac{1}{2} E(k,t) + \frac{1}{4} k^2 \frac{\partial^2 E(k,t)}{\partial k^2} - \frac{1}{4} \gamma k \frac{\partial E(k,t)}{\partial k} - \frac{1}{4} \gamma E(k,t) \right] \delta(\bm 0) \,.
  \label{eq:gipb_advection_term}
\end{align}
Inserting \eqref{eq:gipb_force}, \eqref{eq:gipb_gamma_term} and \eqref{eq:gipb_advection_term} into \eqref{eq:covariance_evolution} yields
\begin{align}
  \frac{\partial E(k,t)}{\partial t} = -2 \nu k^2 E(k,t) + Q(k) &+ \frac{\beta}{4} \gamma \left[ 7 \gamma E(k,t) - 2 k \frac{\partial E(k,t)}{\partial k} +  E(k,t) \right] + \frac{\beta}{4} \left[ - 2 E(k,t) + k^2 \frac{\partial^2 E(k,t)}{\partial k^2} \right] \\
  = -2 \nu k^2 E(k,t) + Q(k) &+ \frac{\beta}{4} \gamma \left[ (7 \gamma+1) E(k,t) - 2 k \frac{\partial E(k,t)}{\partial k} \right] - \frac{\beta}{4} \frac{\partial}{\partial k} \left[ 2 k E(k,t) - k^2 \frac{\partial E(k,t)}{\partial k} \right] \\
  = -2 \nu k^2 E(k,t) + Q(k) &+ \frac{\beta}{4} \gamma (7 \gamma+3) E(k,t) - \frac{\beta}{4} \frac{\partial}{\partial k} \left[ 2 k (1+\gamma) E(k,t) - k^2 \frac{\partial E(k,t)}{\partial k} \right] \label{app_eq:spectral_energy_budget_flux_version}
\end{align}

\section{Solutions and stability of the spectral energy budget equation for general $\gamma$}
\label{sec:spectral_budget_general_gamma}

In the main text, we discussed the $\gamma = 0$ case, for which we obtained a steady state solution. However, a stable stationary solution does not exist for all values of $\gamma$. In this Appendix, we first consider the existence of stationary power-law solutions outside of the forcing and viscous ranges. Then, we search for stationary solutions in the full range of scales. Finally, we discuss the general stability of such stationary solutions.

\begin{figure}
  \includegraphics[width=0.4\textwidth]{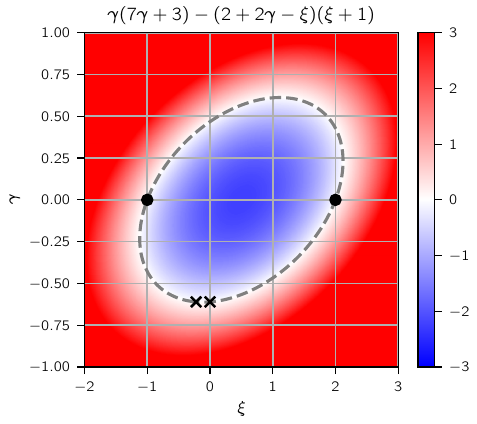}
  \caption{Growth rate of the amplitude of power-law solutions, see \eqref{eq:amplitude_evolution}. Red corresponds to positive growth rates, blue to negative ones. The dashed line indicates zero growth rates which corresponds to stationary solutions. For $|\gamma| > \sqrt{3/8}$ no stationary solutions exist. The black dots illustrate the $k^{-1}$ and $k^2$ solutions for $\gamma=0$ discussed in Sec.~\ref{sec:spectral_energy_budget}. The black crosses demarcate the part of the line where both power-law exponents $\xi_{1,2}$ for a given $\gamma$ are negative, leading to a stationary solution that diverges for $k\to 0$.}
  \label{fig:growth_rate_alpha_gamma}
\end{figure}

It is instructive to seek for self-similar power-law solutions of the form
\begin{equation}
  \label{eq:power_law_solution}
  E(k,t) = A(t) (\lambda k)^{\xi} \, ,
\end{equation}
which are valid in the range of scales where forcing and viscous diffusion are negligible.
Insertion of \eqref{eq:power_law_solution} into \eqref{app_eq:spectral_energy_budget_flux_version} while ignoring the forcing and the viscous terms yields
\begin{equation}
  \frac{4}{\beta} \frac{\dif }{\dif  t} \ln A(t) = \gamma (7\gamma+3) - (2+2\gamma - \xi) (1+\xi)  \, .
  \label{eq:amplitude_evolution}
\end{equation}
The right-hand side specifies the growth rate of the amplitude (non-dimensionalized by $\beta/4$) as a function of $\gamma$ and the exponent $\xi$. Figure~\ref{fig:growth_rate_alpha_gamma} shows this growth rate as a function of $\gamma$ and $\xi$, illustrating the parameter combinations for which stationary states can exist. If a stationary state is achieved, the left-hand side needs to be zero, which implies
\begin{equation}
  \gamma (7\gamma+3) - (2+2\gamma - \xi) (1+\xi) = 0 \,,
\end{equation}
with the solutions
\begin{equation} \label{eq:stationary_power_law_solutions_exponents}
  \xi_{1,2} = \frac{1}{2} + \gamma \pm \frac12 \sqrt{9-24\gamma^2} \,.
\end{equation}
For $\gamma=0$, these are $\xi_1 = -1$ and $\xi_2=2$, as indicated by the dots in figure~\ref{fig:growth_rate_alpha_gamma}. Analogous to the $\gamma=0$ case, we generally expect that the smaller exponent is relevant for wavenumbers beyond the forcing range and the larger exponent is valid for wavenumbers below the forcing range. For $|\gamma|>\sqrt{3/8}$ stationary power-law solutions cease to exist.

To obtain a solution for the full range of scales, we can use the method of Green's functions, analogous to section~\ref{sec:spectral_energy_budget} but for general values of $\gamma$. We seek stationary solutions of~\eqref{eq:spectral_energy_budget} with forcing $Q(k) = \delta(k - k')$. The equation for the Green's function reads
\begin{equation}
  \lambda^2 k^2 G(k,k') - \gamma (7 \gamma+3) G(k,k') + \frac{\partial}{\partial k} \left[ 2 k (1+\gamma) G(k,k') - k^2 \frac{\partial G(k,k')}{\partial k} \right] = \frac{4}{\beta}  \delta(k-k')\, .
\end{equation}
In the interval $\gamma \in [-\sqrt{3/8}, \sqrt{3/8}]$, the solution with appropriate boundary conditions reads
\begin{equation}
  G(k,k') = \frac{4\lambda}{\beta}
  \begin{cases}
    \label{eq:complete_greens_function}
    (\lambda k)^{\frac{1}{2}+\gamma} K_{\sqrt{9-24\gamma^2}/2}(\lambda k) \, (\lambda k')^{-\frac{3}{2}-\gamma} I_{\sqrt{9-24\gamma^2}/2}(\lambda k') & k > k' \\ 
    (\lambda k)^{\frac{1}{2}+\gamma} I_{\sqrt{9-24\gamma^2}/2}(\lambda k) \, (\lambda k')^{-\frac{3}{2}-\gamma} K_{\sqrt{9-24\gamma^2}/2}(\lambda k') & k \le k'  \,, 
   \end{cases}
\end{equation}
where $I_\alpha(x)$ and $K_\alpha(x)$ are the $\alpha$th-order modified Bessel functions of the first and second kind, respectively. Note that this solution diverges at $k\to 0$ in the small interval $\gamma \in [-\sqrt{3/8}, -(1 + \sqrt{57})/14[$, which can be shown using the asymptotics
\begin{align}
  I_\alpha(x) \stackrel{x\to 0}{\sim} \frac{(x/2)^{\alpha }}{\Gamma (\alpha +1)}\,,
\end{align}
where $\Gamma(x)$ is the gamma function. This corresponds to the interval of $\gamma$ values in which both exponents~\eqref{eq:stationary_power_law_solutions_exponents} are negative, as demarcated by black crosses in figure~\ref{fig:growth_rate_alpha_gamma}. Nevertheless, in the full interval $\gamma \in [-\sqrt{3/8}, \sqrt{3/8}]$, solutions constructed using~\eqref{eq:complete_greens_function} satisfy the no-energy-flux boundary conditions,
\begin{align}
  J(k) = \frac{\beta}{4} \left[ 2 k (1+\gamma) E(k,t) - k^2 \frac{\partial E(k,t)}{\partial k} \right]
  \xrightarrow{k \to 0, \infty} 0\,,
\end{align}
provided that $Q(k)$ converges sufficiently fast to zero for $k \to 0, \infty$.
It can be checked that the expression~\eqref{eq:complete_greens_function} reduces to the Green's function given in section~\ref{sec:spectral_energy_budget} when $\gamma=0$.

We now turn to the problem of stability. From the kinematic dynamo problem (corresponding to the case $\gamma = 1$), it is well known that passive vector models of this type may grow without bounds~\cite{AdzhemyanE2001,VincenziJoSP2002,ArponenPRE2010}. Since for our case of spatially linear advecting field, we have derived the explicit evolution equation~\eqref{app_eq:spectral_energy_budget_flux_version} for the energy spectrum, we can assess stability directly based on this equation. Mathematically, the question of stability of the equation relates to the spectrum (in the sense of functional analysis) of the right-hand side differential operator. Here, we provide a more intuitive analysis of the stability.

Any perturbation around a potential stationary state satisfies the homogeneous (i.e., unforced) version of equation~\eqref{app_eq:spectral_energy_budget_flux_version}. If we can find permissible eigenfunctions of the linear operator with positive eigenvalues, then this indicates that solutions will be unstable. The eigenfunction equation reads
\begin{align}
  \sigma E(k)
  = -2 \nu k^2 E(k) &\,+\, \frac{\beta}{4} \gamma (7 \gamma+3) E(k) - \frac{\beta}{4} \dod{}{k} \left[ 2 k (1+\gamma) E(k) - k^2 \dod{E(k)}{k} \right]
\end{align}
with eigenvalue $\sigma$. By substituting $E(k) = k^{\gamma+1/2}$ $f(\lambda k)$ and $x=\lambda k$, we can rewrite this equation as a modified Bessel equation,
\begin{align}
    x^2 f''(x) + x f'(x) - \left[x^2 + \alpha^2(\gamma, \sigma) \right] f(x) &= 0 \,,
\end{align}
where
\begin{align}
  \alpha^2(\gamma, \sigma) = \frac{4}{\beta} \sigma - \frac{24\gamma^2-9}{4}
\end{align}
is the squared Bessel order. This equation is solved by the modified Bessel functions $I_\alpha(x)$ and $K_\alpha(x)$. The set of permissible eigenfunctions depends on the specifics of the boundary conditions. For $x \to \infty$, all modified Bessel functions of the first kind $I_\alpha$ grow exponentially, which excludes them under all reasonable boundary conditions. For the modified Bessel functions of the second kind $K_\alpha$, the case is more complicated. While real-order $K_\alpha$ diverge at zero, imaginary-order $K_\alpha$ remain bounded, and can thus likely be used to parameterize perturbations around the stationary state. If any of them correspond to positive eigenvalues $\sigma$, then the equation is expected to be unstable. The order $\alpha(\gamma, \sigma)$ turns imaginary if
\begin{align}
  \frac{4}{\beta} \sigma &< \frac{24\gamma^2-9}{4}
\end{align}
and so there exist permissible eigenfunctions with eigenvalue $\sigma > 0$ if
\begin{align}
  \frac{24\gamma^2-9}{4} &> 0 \\
  \Leftrightarrow \gamma &\notin [-\sqrt{3/8}, \sqrt{3/8}] \,.
\end{align}
This implies that at least outside the interval $\gamma \in [-\sqrt{3/8}, \sqrt{3/8}]$, solutions are expected to be unstable. This is consistent with the range in which we found stationary power law solutions. This also means that solutions are unstable for $\gamma = \pm 1$, which is consistent with the literature findings for sufficiently smooth advecting fields~\cite{AdzhemyanE2001,VincenziJoSP2002,ArponenJSP2007,ArponenPRE2010}.

\section{Derivation of the evolution equation for the characteristic functional}
\label{app:characteristic_functional_equation}

In this section, we derive the Hopf equation for the passive vector field in Fourier space, both unconditional and conditional on the realization of the advecting velocity field. For the conditional equation, we perform averages only over the realizations of the forcing and potentially over initial conditions. Taking the time derivative of the conditional characteristic functional~\eqref{eq:def_cond_cf_fourier} and inserting Eq.~\eqref{eq:passive_vector_model_fourier} results in
\begin{align}
  \label{eq:hopf_fourier_space_starting_point}
  \partial_t \Phi^{\bm v}[\bm \theta](t) &= \mathrm{i} \int \dif \bm k \, \hat{\theta}_i(-\bm k) \cdot \left\langle \left[ \partial_t \hat{u}_i(\bm k,t) \right] \exp \left[ \mathrm{i} \int \dif \bm k' \, \hat{\bm \theta}(-\bm k') \cdot  \hat{\bm u}(\bm k',t) \right] \right\rangle_{\!\!\bm{F}} \\
  &= \mathrm{i} \int \dif  \bm k \, \hat{\theta}_i(-\bm k) \bigg\langle \Big[ \begin{aligned}[t] &\big[- \mathrm i k_l V_l(t) - \nu k^2 \big] \hat{u}_i(\bm k,t)  \\
  &+ P_{ij}(\bm k) \left[ k_l B_{lm}(t) \partial_m \hat{u}_j(\bm k,t)+ \gamma B_{jm}(t) \hat{u}_m(\bm k,t) \right] \nonumber \\
  &+ \hat{F}_i(\bm k,t) \Big] \exp \left[ \mathrm{i} \int \dif  \bm k' \, \hat{\bm \theta}(-\bm k') \cdot  \hat{\bm u}(\bm k',t)\right] \bigg\rangle_{\!\!\bm{F}} \, .  \end{aligned}  \nonumber  
\end{align}
The simplest term to evaluate is the viscous diffusion term, which we can rewrite using the Hopf formalism \cite{hopf1952,Monin2013BothVolumes}:
\begin{align}
  \left\langle - \nu k^2 \hat{u}_i(\bm k,t) \exp \left[ \mathrm{i} \int \dif  \bm k' \, \hat{\bm \theta}(-\bm k') \cdot  \hat{\bm u}(\bm k',t) \right] \right\rangle_{\!\! \bm F} = \mathrm{i} \nu k^2 \frac{\delta}{\delta \hat{\theta}_i(-\bm k)} \Phi^{\bm v}[\bm \theta](t) \,.
\end{align}
For the forcing term, we make use of Gaussian integration by parts \cite{novikov1965,FurutsuJRNBSD1963,DonskerAFS1964}. As part of these calculations, we also need to evaluate the mean response function, for which we require the formal solution of Eq.~\eqref{eq:passive_vector_model_fourier} given by
\begin{equation}
  \hat{u}_i(\bm k,t)  =  \hat{u}_i(\bm k,0) + \int_0^t \dif  s \, \left[-\left[\mathrm i k_l V_l(s) + \nu k^2 \right] \hat{u}_i(\bm k,s) + P_{ij}(\bm k) \left[ k_l B_{lm}(s) \partial_m \hat{u}_j(\bm k,s)+ \gamma B_{jm}(s) \hat{u}_m(\bm k,s) \right] + \hat{F}_i(\bm k,s)\right] \, .
\end{equation}
The forcing contribution amounts to the classic problem of randomly forced turbulence \cite{novikov1965}:
\begin{align}
  \bigg\langle \hat{F}_i(\bm k,t) \exp \left[ \mathrm{i} \int \dif  \bm k' \, \hat{\bm \theta}(-\bm k') \cdot  \hat{\bm u}(\bm k',t) \right] \bigg\rangle_{\!\! \bm F} &= \psi_{ij}(\bm k) \bigg\langle \frac{\delta}{\delta \hat{F}_j(-\bm k,t)} \exp \left[ \mathrm{i} \int \dif  \bm k' \, \hat{\bm \theta}(-\bm k') \cdot  \hat{\bm u}(\bm k',t) \right] \bigg\rangle_{\!\! \bm F} \\
  &= \frac{i}{2} \psi_{ij}(\bm k) \, \hat{\theta}_j(\bm k,t) \, \Phi^{\bm v}[\bm \theta](t) \,.
\end{align}
The first equality is obtained by applying Gaussian integration by parts with respect to the spatially homogeneous additive forcing with spectral energy tensor $\psi_{ij}(\bm k)$, and the second equality results from the evaluation of the mean response function. As long as we average only over the forcing, the advecting velocity field can be treated as deterministic. As a result, the corresponding terms can be evaluated in the same way as the viscous term, leading to the conditional Hopf equation
\begin{align} \label{eq:conditional_hopf_equation_general_gamma}
  \partial_t \Phi^{\bm v}[\bm \theta](t) = \int \dif \bm k \, \hat{\theta}_i(-\bm k) \Biggl(
  &\left[-\mathrm{i}k_l V_l(t) - \nu k^2\right] \frac{\delta}{\delta \hat{\theta}_i(-\bm k)} - \frac{1}{2} \psi_{ij}(\bm k) \, \hat{\theta}_j(\bm k) \\
  &+ P_{ij}(\bm k) \left[ k_l B_{lm}(t) \partial_m \frac{\delta}{\delta \hat{\theta}_j(-\bm k)}+ \gamma B_{jm}(t)  \frac{\delta}{\delta \hat{\theta}_m(-\bm k)} \right]  \Biggr) \Phi^{\bm v}[\bm \theta](t) \, .  \nonumber
\end{align}

For the unconditional Hopf equation, we additionally need to average over the advecting velocity field. The viscous and forcing terms keep the same form. The computation of the random sweeping term follows an analogous but more involved computation using Gaussian integration by parts for $\bm v$. It results in~
\begin{align}
  - \mathrm i k_l \frac{\delta}{\delta \hat{\theta}_i(-\bm k)} \left\langle V_l(t) \Phi^{\bm v}[\bm \theta](t) \right\rangle_{\bm v}
  &= -\frac{\alpha}{2} k_l \frac{\delta}{\delta \hat{\theta}_i(-\bm k)} \int \dif \bm k'\, \hat{\theta}_m(-\bm k') \, k'_l \, \frac{\delta}{\delta \hat{\theta}_m(-\bm k')} \Phi[\bm \theta](t) \, .
\end{align}
For the terms related to the stretching, we need to evaluate
\begin{align}
  \left\langle B_{lm}(t) \Phi^{\bm v}[\bm \theta](t) \right\rangle_{\bm v} 
  &= \frac{1}{2} D_{lmab} \int \dif  \bm k' \, \hat{\theta}_c(-\bm k') P_{cd}(\bm k') \left[ k'_a \partial'_b \frac{\delta}{\delta \hat{\theta}_d(-\bm k')} + \gamma \delta_{ad} \frac{\delta}{\delta \hat{\theta}_b(-\bm k')} \right] \Phi[\bm \theta](t) \,.
\end{align}
The fully averaged Hopf equation reads
\begin{align} \label{eq:full_hopf_equation_general_gamma}
  \partial_t \Phi[\bm \theta](t) = \int \dif \bm k \, \hat{\theta}_i(-\bm k) \Biggl(
  &- \nu k^2 \frac{\delta}{\delta \hat{\theta}_i(-\bm k)} - \frac{1}{2} \psi_{ij}(\bm k) \, \hat{\theta}_j(\bm k)
  -\frac{\alpha}{2} k_l \frac{\delta}{\delta \hat{\theta}_i(-\bm k)} \int \dif \bm k'\, \hat{\theta}_m(-\bm k') \, k'_l \, \frac{\delta}{\delta \hat{\theta}_m(-\bm k')} \\
  &+ \frac{1}{2} D_{lmab} \int \dif  \bm k' \, \hat{\theta}_c(-\bm k') P_{ij}(\bm k) \left[ k_l \partial_m \frac{\delta}{\delta \hat{\theta}_j(-\bm k)}+ \gamma \delta_{jl} \frac{\delta}{\delta \hat{\theta}_m(-\bm k)} \right] \nonumber\\
  &\qquad\qquad\qquad\times P_{cd}(\bm k') \left[ k'_a \partial'_b \frac{\delta}{\delta \hat{\theta}_d(-\bm k')} + \gamma \delta_{ad} \frac{\delta}{\delta \hat{\theta}_b(-\bm k')} \right]\Biggr) \Phi[\bm \theta](t) \, ,  \nonumber
\end{align}
or, in the case $\gamma = 0$,
\begin{align}
  \partial_t \Phi[\bm \theta](t) = \int \dif \bm k \, \hat{\theta}_i(-\bm k) \Biggl(
  &- \nu k^2 \frac{\delta}{\delta \hat{\theta}_i(-\bm k)} - \frac{1}{2} \psi_{ij}(\bm k) \, \hat{\theta}_j(\bm k)
  -\frac{\alpha}{2} k_l \frac{\delta}{\delta \hat{\theta}_i(-\bm k)} \int \dif \bm k'\, \hat{\theta}_m(-\bm k') \, k'_l \, \frac{\delta}{\delta \hat{\theta}_m(-\bm k')}\\
  &+ \frac{1}{2} D_{lmab} \int \dif  \bm k' \, \hat{\theta}_c(-\bm k') P_{ij}(\bm k) k_l \partial_m \frac{\delta}{\delta \hat{\theta}_j(-\bm k)}  P_{cd}(\bm k')  k'_a \partial'_b \frac{\delta}{\delta \hat{\theta}_d(-\bm k')} \Biggr) \Phi[\bm \theta](t) \, . \nonumber
\end{align}
The symmetry properties of these equations are analyzed in Appendix~\ref{sec:hopf_symmetries}.

\section{Symmetries of the Hopf equations}
\label{sec:hopf_symmetries}
Although the advecting field $\bm v(\bm x, t)$ is statistically inhomogeneous, we find that the single-time statistics of the advected field $\bm u(\bm x, t)$ are statistically homogeneous and isotropic (given suitable initial conditions). We also find that the single-time statistics of the sub-ensembles conditional on the time series of $\bm v(\bm x, t)$ are anisotropic but homogeneous. The intuitive explanation for homogeneity of single-time statistics being preserved is that the gradients of the advecting field are spatially uniform. Here, we show these symmetry properties based on the Hopf equations derived in Appendix~\ref{app:characteristic_functional_equation}.

Statistical homogeneity can be defined as the invariance of statistical averages under translations of space $\tilde{\bm u}(\bm x, t) = \bm u(\bm x + \bm r, t)$ for $\bm r \in \mathbb{R}^3$. Translated into Fourier space, the transformation reads $\tilde{\hat{\bm u}}(\bm k, t) = e^{\mathrm{i} \bm k\cdot \bm r} \hat{\bm u}(\bm k, t)$. The characteristic functional $\Phi[\bm \theta]$~\eqref{eq:def_full_cf_fourier} contains the full single-time statistics of the model. Hence single-time statistics are homogeneous if and only if for all $\bm r\in\mathbb{R}^3$
\begin{align}
  \Phi\left[\hat{\bm \eta}(\bm k) = e^{-\mathrm{i}\bm k\cdot \bm r}\hat{\bm \theta}(\bm k)\right]
  &= \left\langle \exp \left[ \mathrm{i} \int \dif \bm k \,\hat{\bm \theta}(-\bm k) \cdot e^{\mathrm{i}\bm k\cdot \bm r} \hat{\bm u}(\bm k,t) \right] \right\rangle \\
  &= \left\langle \exp \left[ \mathrm{i} \int \dif \bm k \, \hat{\bm \theta}(-\bm k) \cdot \hat{\bm u}(\bm k,t) \right] \right\rangle \\
  &= \Phi[\bm \theta] \,.
\end{align}
In particular, for functionals satisfying this property we can take the derivative with respect to $r_i$, evaluate at $\bm r=\bm 0$, and obtain
\begin{align} \label{eq:functional_homogeneity_constraint}
  0 = \eval{\dpd{}{r_i} \Phi[\bm \theta]}_{\bm r = \bm 0}
  = \eval{\dpd{}{r_i} \Phi\left[\hat{\bm \eta}(\bm k) = e^{-\mathrm{i}\bm k\cdot \bm r}\hat{\bm \theta}(\bm k)\right]}_{\bm r = \bm 0}
  = -\mathrm{i}\int \dif \bm k\, k_i \hat{\theta}_j(\bm k) \frac{\delta}{\delta \hat{\theta}_j(\bm k)} \Phi[\bm \theta] \,,
\end{align}
which will be used below.
Statistical isotropy can be defined as the invariance of statistical averages under rotations of space $\tilde{\bm u}(\bm x, t) = O^{T}\bm u(O\bm x, t)$ for $O\in\mathrm{SO}(3)$. Translated into Fourier space, this reads $\tilde{\hat{\bm u}}(\bm k, t) = O^{T} \hat{\bm u}(O\bm k, t)$. Using the characteristic functional, we thus find that single-time statistics are isotropic if and only if for all $O \in \mathrm{SO}(3)$
\begin{align}
  \Phi\left[\hat{\bm \eta}(\bm k) = O \hat{\bm \theta}(O^{T}\bm k)\right]
  &= \left\langle \exp \left[ \mathrm{i} \int \dif \bm k \, \hat{\bm \theta}(-\bm k) \cdot O^{T} \hat{\bm u}(O\bm k,t) \right] \right\rangle \\
  &= \left\langle \exp \left[ \mathrm{i} \int \dif \bm k \, \hat{\bm \theta}(-\bm k) \cdot \hat{\bm u}(\bm k,t) \right] \right\rangle \\
  &= \Phi[\bm \theta] \,.
\end{align}

We proceed by analyzing the symmetry properties of the conditional Hopf equation~\eqref{eq:conditional_hopf_equation_general_gamma}. Assuming $\Phi^{\bm v}[\bm \theta](t)$ is initially homogeneous, we apply the transformation $\hat{\bm \theta}(\bm k) = e^{\mathrm{i}\bm k\cdot \bm r}\hat{\bm \eta}(\bm k)$ to the right-hand side of~\eqref{eq:conditional_hopf_equation_general_gamma}. If it remains invariant, then the equation preserves homogeneity. To transform the functional derivatives, by chain rule, we observe for some functional $F$ that
\begin{align}
  \frac{\delta}{\delta \hat{\theta}_j(-\bm k)} F\left[\hat{\bm \eta}(\bm k') = e^{-\mathrm{i}\bm k'\cdot\bm r}\hat{\bm \theta}(\bm k')\right]
  &= \int \dif\bm k'\, e^{-\mathrm{i}\bm k'\cdot\bm r} \delta_{jk} \delta(\bm k' + \bm k) \frac{\delta}{\delta \hat{\eta}_k(\bm k')} F[\bm \eta] = e^{\mathrm{i}\bm k\cdot\bm r}\frac{\delta}{\delta \hat{\eta}_j(-\bm k)} F[\bm \eta]\,.
\end{align}
Thus, we get invariance of terms of the form
\begin{align} \label{eq:invariant_term1}
  \hat{\theta}_i(-\bm k) \hat{\theta}_j(\bm k) &= e^{-\mathrm{i}\bm k\cdot\bm r}\hat{\eta}_i(-\bm k) e^{\mathrm{i}\bm k\cdot\bm r}\hat{\eta}_j(\bm k) = \hat{\eta}_i(-\bm k) \hat{\eta}_j(\bm k)
\end{align}
as well as
\begin{align} \label{eq:invariant_term2}
\hat{\theta}_i(-\bm k) \frac{\delta}{\delta \hat{\theta}_j(-\bm k)}
&= e^{-\mathrm{i}\bm k\cdot\bm r}\hat{\eta}_i(-\bm k) e^{\mathrm{i}\bm k\cdot\bm r}\frac{\delta}{\delta \hat{\eta}_j(-\bm k)}
= \hat{\eta}_i(-\bm k) \frac{\delta}{\delta \hat{\eta}_j(-\bm k)} \,.
\end{align}
One term of a different form remains, which contains a $\bm k$-derivative. However, it can also be shown to be invariant:
\begin{align}
  &\quad\int \dif \bm k \, \hat{\theta}_i(-\bm k) P_{ij}(\bm k) k_l B_{lm}(t) \partial_m \frac{\delta}{\delta \hat{\theta}_j(-\bm k)} \Phi^{\bm v}[\bm \theta](t) \\
  &= \int \dif \bm k \, e^{-\mathrm{i}\bm k\cdot\bm r} \hat{\eta}_i(-\bm k) P_{ij}(\bm k) k_l B_{lm}(t) \partial_m \left[ e^{\mathrm{i}\bm k\cdot\bm r}\frac{\delta}{\delta \hat{\eta}_j(-\bm k)} \Phi^{\bm v}[\bm \eta](t)\right] \\
  &= \int \dif \bm k \, \hat{\eta}_i(-\bm k) P_{ij}(\bm k) k_l B_{lm}(t) \left(ir_m + \partial_m\right)\frac{\delta}{\delta \hat{\eta}_j(-\bm k)} \Phi^{\bm v}[\bm \eta](t) \\
  &= \int \dif \bm k \, \hat{\eta}_i(-\bm k) P_{ij}(\bm k) k_l B_{lm}(t) \partial_m \frac{\delta}{\delta \hat{\eta}_j(-\bm k)} \Phi^{\bm v}[\bm \eta](t)\,.
\end{align}
In the first step, we inserted the transformation, leading to a product rule in the second step. In the last step, we used incompressibility,
\begin{align}
  P_{ij}(\bm k) \frac{\delta}{\delta \hat{\eta}_j(-\bm k)} \Phi[\bm \eta](t)
  = \frac{\delta}{\delta \hat{\eta}_i(-\bm k)} \Phi[\bm \eta](t)\,,
\end{align}
to identify a term of the form~\eqref{eq:functional_homogeneity_constraint}, which is zero by homogeneity of the functional itself. This shows that single-time statistics of the sub-ensembles remain homogeneous given statistically homogeneous initial conditions. Additional averaging does not break this symmetry, so the full ensemble in~\eqref{eq:full_hopf_equation_general_gamma} must also preserve homogeneity.

The same strategy can be used to show isotropy. If applying the transformation $\hat{\bm \theta}(\bm k) = O^T \hat{\bm \eta}(O\bm k)$ leaves the form of the right-hand side invariant (while assuming isotropy of the characteristic functional itself), then the equation preserves isotropy. For the conditional Hopf equation~\eqref{eq:conditional_hopf_equation_general_gamma}, the terms with $\mathrm{B}(t)$ prevent invariance. For the full Hopf equation~\eqref{eq:full_hopf_equation_general_gamma}, however, we can show that the right-hand side remains invariant. To this end, transform the functional derivatives by chain rule, again:
\begin{align}
  \frac{\delta}{\delta \hat{\theta}_i(-\bm k)} F[\hat{\bm \eta}(\bm k') = \mathrm{O}\hat{\bm \theta}(\mathrm{O}^T\bm k')]
  &= \int \dif\bm k'\, O_{jk} \delta_{ik} \delta(\bm k + \mathrm{O}^T\bm k') \frac{\delta}{\delta \hat{\eta}_j(\bm k')} F[\bm \eta] \\
  &= O^T_{ij} \frac{\delta}{\delta \hat{\eta}_j(-\mathrm{O}\bm k)} F[\bm \eta]\,,
\end{align}
where we performed an integral substitution $\bm k' \to \mathrm{O}\bm k'$ before evaluating the delta function.
The viscous term remains invariant by
\begin{align}
  \int \dif \bm k \, \hat{\theta}_i(-\bm k) \nu k^2 \frac{\delta}{\delta \hat{\theta}_i(-\bm k)}
  &= \int \dif \bm k \, O^T_{ij} \hat{\eta}_j(-O\bm k) \nu k^2 O^T_{ik} \frac{\delta}{\delta \hat{\eta}_k(-O\bm k)}\\
  &= \int \dif \bm k \, \hat{\eta}_i(-\bm k) \nu k^2 \frac{\delta}{\delta \hat{\eta}_i(-\bm k)} \,.
\end{align}
For the forcing, we have
\begin{align}
  \int \dif \bm k \, \hat{\theta}_i(-\bm k) \psi_{ij}(\bm k) \hat{\theta}_j(\bm k)
  &= \int \dif \bm k \, O^T_{ik}\hat{\eta}_k(-O\bm k) \psi_{ij}(\bm k) O^T_{jl}\hat{\eta}_l(O\bm k) \\
  &= \int \dif \bm k \, \hat{\eta}_k(-\bm k) \left(O_{ki} \psi_{ij}(O^T\bm k) O^T_{jl}\right) \hat{\eta}_l(\bm k)\\
  &= \int \dif \bm k \, \hat{\eta}_k(-\bm k) \psi_{kl}(\bm k) \hat{\eta}_l(\bm k) \,,
\end{align}
by the isotropic form of the forcing spectrum~\eqref{eq:forcing_spectrum}. For the sweeping term, we have 
\begin{align}
  &\quad \int \dif \bm k \, \hat{\theta}_i(-\bm k) \frac{\alpha}{2} k_l \frac{\delta}{\delta \hat{\theta}_i(-\bm k)} \int \dif \bm k'\, \hat{\theta}_m(-\bm k') \, k'_l \, \frac{\delta}{\delta \hat{\theta}_m(-\bm k')} \Phi[\bm \theta](t)\nonumber\\
  &= \int \dif \bm k \, (O^T \hat{\bm \eta})_i(-O\bm k) \frac{\alpha}{2} k_l \frac{\delta}{\delta (O^T \hat{\bm \eta})_i(-O\bm k)} \int \dif \bm k'\, (O^T \hat{\bm \eta})_m(-O\bm k') \, k'_l \, \frac{\delta}{\delta (O^T \hat{\bm \eta})_m(-O\bm k')} \Phi[\bm \eta](t)\\
  &= \int \dif \bm k \, \hat{\eta}_i(-\bm k) \frac{\alpha}{2} k_l \frac{\delta}{\delta \hat{\eta}_i(-\bm k)} \int \dif \bm k'\, \hat{\eta}_m(-\bm k') \, k'_l \, \frac{\delta}{\delta \hat{\eta}_m(-\bm k')} \Phi[\bm \eta](t) \,,
\end{align}
where we substituted $\bm k \rightarrow \mathrm{O}^T \bm k$ and $\bm k' \rightarrow \mathrm{O}^T \bm k'$.
For computing the gradient-related term, we use the same substitutions (in the second step of the following computation), which also imply $\partial_m \rightarrow O^T_{mn} \partial_n$ and $\partial_b' \rightarrow O^T_{bc} \partial_c'$:
\begin{align}
  &\quad D_{lmab} \int \dif \bm k \, \hat{\theta}_i(-\bm k) P_{ij}(\bm k) \int \dif  \bm k' \, \hat{\theta}_c(-\bm k') P_{cd}(\bm k') \nonumber\\
  &\quad\quad\times \left[ k_l \partial_m \frac{\delta}{\delta \hat{\theta}_j(-\bm k)}+ \gamma \delta_{jl} \frac{\delta}{\delta \hat{\theta}_m(-\bm k)} \right]  \left[ k'_a \partial'_b \frac{\delta}{\delta \hat{\theta}_d(-\bm k')} + \gamma \delta_{ad} \frac{\delta}{\delta \hat{\theta}_b(-\bm k')} \right] \Phi[\bm \theta](t) \nonumber \\
  &= D_{lmab} \int \dif \bm k \, (O^T \hat{\bm \eta})_i(-O\bm k) P_{ij}(\bm k) \int \dif  \bm k' \, (O^T \hat{\bm \eta})_c(-O\bm k') P_{cd}(\bm k') \nonumber\\
  &\quad\quad\times \left[ k_l \partial_m \frac{\delta}{\delta (O^T \hat{\bm \eta})_j(-O\bm k)}+ \gamma \delta_{jl} \frac{\delta}{\delta (O^T \hat{\bm \eta})_m(-O\bm k)} \right]  \left[ k'_a \partial'_b \frac{\delta}{\delta (O^T \hat{\bm \eta})_d(-O\bm k')} + \gamma \delta_{ad} \frac{\delta}{\delta (O^T \hat{\bm \eta})_b(-O\bm k')} \right] \Phi[\bm \eta](t) \\
  &= D_{lmab} O^T_{lp} O^T_{mn} O^T_{aq} O^T_{br} \int \dif \bm k \, \hat{\eta}_i(-\bm k) (\mathrm{O}\mathrm{P}\mathrm{O}^T)_{ij}(\mathrm{O}^T\bm k) \int \dif  \bm k' \, \hat{\eta}_c(-\bm k') (\mathrm{O}\mathrm{P}\mathrm{O}^T)_{cd}(\mathrm{O}^T\bm k') \nonumber\\
  &\quad\quad\times \left[ k_p \partial_n \frac{\delta}{\delta \hat{\eta}_j(-\bm k)}+ \gamma \delta_{jp} \frac{\delta}{\delta \hat{\eta}_n(-\bm k)} \right]  \left[ k'_q \partial'_r \frac{\delta}{\delta \hat{\eta}_d(-\bm k')} + \gamma \delta_{qd} \frac{\delta}{\delta \hat{\eta}_r(-\bm k')} \right] \Phi[\bm \eta](t) \\
  &= D_{lmab} \int \dif \bm k \, \hat{\eta}_i(-\bm k) P_{ij}(\bm k) \int \dif  \bm k' \, \hat{\eta}_c(-\bm k') P_{cd}(\bm k') \nonumber\\
  &\quad\quad\times \left[ k_l \partial_m \frac{\delta}{\delta \hat{\eta}_j(-\bm k)}+ \gamma \delta_{jl} \frac{\delta}{\delta \hat{\eta}_m(-\bm k)} \right]  \left[ k'_a \partial'_b \frac{\delta}{\delta \hat{\eta}_d(-\bm k')} + \gamma \delta_{ad} \frac{\delta}{\delta \hat{\eta}_b(-\bm k')} \right] \Phi[\bm \eta](t) \,.
\end{align}
In the last step, we used
\begin{align}
  D_{lmab} O^T_{lp} O^T_{mn} O^T_{aq} O^T_{br} =  D_{pnqr}
\end{align}
and
\begin{align}
  (\mathrm{O}\mathrm{P}\mathrm{O}^T)_{ij}(\mathrm{O}^T\bm k) &= O_{ik}\delta_{kl}O^T_{lj} - \frac{1}{k^2} O_{ik}O^T_{km}k_m O^T_{ln}k_n O^T_{lj}
  = \delta_{ij} - \frac{k_i k_j}{k^2} = P_{ij}(\bm k) \,.
\end{align}
Therefore, the full Hopf equation conserves isotropy.

\section{Gaussian solution to conditional Hopf equation for general $\gamma$}
\label{sec:gaussian_solution_conditional_hopf_general_gamma}
The result that the conditional Hopf equation is solved by a Gaussian characteristic functional as shown in section~\ref{sec:stat_field_theory} can be generalized to $\gamma \neq 0$. To this end, insert the Gaussian ansatz~\eqref{eq:gauss_char_func} into the general conditional Hopf equation~\eqref{eq:conditional_hopf_equation_general_gamma}, yielding 
\begin{align}
  -\frac{1}{2} \int\dif\bm k\, \hat{\theta}_i(-\bm k) \partial_t \phi^{\bm v}_{ij}(\bm k, t)\hat{\theta}_j(\bm k)
  &= -\int\dif\bm k\, \hat{\theta}_i(-\bm k) \left(\left[-\mathrm{i}k_l V_l(t) - \nu k^2 \right] \phi^{\bm v}_{ij}(\bm k, t) + \frac{1}{2} \psi_{ij}(\bm k) \right)\hat{\theta}_j(\bm k) \\
  &\quad - \int\dif\bm k\, \hat{\theta}_i(-\bm k)  P_{ip}(\bm k)  k_l B_{lm}(t) \partial_m \left(\phi^{\bm v}_{pj}(\bm k, t)\hat{\theta}_j(\bm k) \right) \nonumber \\
  &\quad - \int\dif\bm k\, \hat{\theta}_i(-\bm k) \gamma P_{il}(\bm k) B_{lm}(t) \phi^{\bm v}_{mj}(\bm k, t) \hat{\theta}_{j}(\bm k)  \nonumber \\
  &= -\int\dif\bm k\, \hat{\theta}_i(-\bm k) \left(- \nu k^2 \phi^{\bm v}_{ij}(\bm k, t) + \frac{1}{2} \psi_{ij}(\bm k) \right)\hat{\theta}_j(\bm k) \\
  &\quad - \frac12 \int\dif\bm k\, \hat{\theta}_i(-\bm k)  P_{ip}(\bm k) P_{jq}(\bm k)  k_l B_{lm}(t) \left(\partial_m \phi^{\bm v}_{pq}(\bm k, t) \right)\hat{\theta}_j(\bm k) \nonumber \\
  &\quad - \frac12 \int\dif\bm k\, \hat{\theta}_i(-\bm k) \gamma \left[ P_{il}(\bm k) B_{lm}(t) \phi^{\bm v}_{mj}(\bm k, t) +  P_{jl}(\bm k) B_{lm}(t) \phi^{\bm v}_{im}(\bm k, t) \right] \hat{\theta}_{j}(\bm k)\,,  \nonumber
\end{align}
where we rewrote the first terms as in section~\ref{sec:stat_field_theory} and the last term such that it is invariant under the symmetry $\phi^{\bm v}_{ij}(\bm k, t) = \phi^{\bm v}_{ji}(-\bm k, t)$. Finally, we again isolate the equation of integrands
\begin{align} \label{eq:subensemble_spectrum_evo_general_gamma}
  \partial_t \phi^{\bm v}_{ij}(\bm k, t) &= -2 \nu k^2 \phi^{\bm v}_{ij}(\bm k, t) + P_{ip}(\bm k) P_{jq}(\bm k) k_l B_{lm}(t) \partial_m \phi^{\bm v}_{pq}(\bm k, t) \\
  &\quad + \gamma P_{il}(\bm k) B_{lm}(t) \phi^{\bm v}_{mj}(\bm k, t) + \gamma P_{jl}(\bm k) B_{lm}(t) \phi^{\bm v}_{im}(\bm k, t) + \psi_{ij}(\bm k) \,. \nonumber
\end{align}
By construction, any solution to~\eqref{eq:subensemble_spectrum_evo_general_gamma} corresponds to a solution of the conditional Hopf equation~\eqref{eq:conditional_hopf_equation_general_gamma} through the ansatz~\eqref{eq:gauss_char_func}. Solutions to the full Hopf equation~\eqref{eq:full_hopf_equation_general_gamma} can be constructed by averaging these Gaussian functionals over the fluctuating spectrum tensors $\phi^{\bm v}_{ij}(\bm k, t)$.

\section{Numerical method for general $\gamma$}
\label{sec:numerics_general_gamma}
For the case $\gamma \neq 0$, the evolution equation~\eqref{eq:subensemble_spectrum_evo_general_gamma} for the sub-ensemble spectrum tensor was derived in the previous section.
Fully analogously to the case $\gamma = 0$, the $\bm k$-derivative can be removed by method of characteristics (compare eq.~\eqref{eq:subensemble_spectrum_characteristic_evo}), leading to the equation
\begin{align}
  \partial_t \tilde{\phi}^{\bm v}_{ij}(\bm q, t) &= - 2 \nu \tilde{k}^2 \tilde{\phi}^{\bm v}_{ij}(\bm q, t)
  + \frac{\tilde{k}_i \tilde{k}_l}{\tilde{k}^2} B_{lm}(t) \tilde{\phi}^{\bm v}_{mj}(\bm q, t)
  + \frac{\tilde{k}_j \tilde{k}_l}{\tilde{k}^2} B_{lm}(t) \tilde{\phi}^{\bm v}_{im}(\bm q, t)\\
  &\quad +\gamma P_{il}(\tilde{\bm k}) B_{lm}(t) \tilde{\phi}^{\bm v}_{mj}(\bm q, t) + \gamma P_{jl}(\tilde{\bm k}) B_{lm}(t) \tilde{\phi}^{\bm v}_{im}(\bm q, t) + \psi_{ij}(\tilde{\bm k}) \,. \nonumber
\end{align}
Due to the additional $\gamma$ terms, the reduction to a scalar field as described at the end of section~\ref{sec:stat_field_theory} is not possible anymore. As a result, simulations need to include the full spectrum tensor, but symmetry allows us to keep only 6 out of the 9 components.

For time stepping, we define a tensorial compensated spectrum (compare eq.~\eqref{eq:def_compensated_spectrum}),
\begin{align}
  Z_{ij}(\bm q, t) &= e^{A(\bm q, t)} \tilde{\phi}^{\bm v}_{ij}(\bm q, t) 
  \,,
\end{align}
and the corresponding evolution equation features more terms:
\begin{align}
  \dpd{}{t} Z_{ij}(\bm q, t) &= \frac{\tilde{k}_i \tilde{k}_l}{\tilde{k}^2} B_{lm}(t) Z_{mj}(\bm q, t)
  + \frac{\tilde{k}_j \tilde{k}_l}{\tilde{k}^2} B_{lm}(t) Z_{im}(\bm q, t) \\
  &\quad
  + \gamma P_{il}(\tilde{\bm k}) B_{lm}(t) Z_{mj}(\bm q, t)
  + \gamma P_{jl}(\tilde{\bm k}) B_{lm}(t) Z_{im}(\bm q, t)
  + e^{A(\bm q, t)} \psi_{ij}(\tilde{\bm k})\,. \nonumber
\end{align}
The numerical method stays the same for the deformation tensor $\mathrm{W}(t)$ and the integrating factor $A(\bm q, t)$, but it needs to be adapted for the compensated spectrum. The intermediate step reads
\begin{align}
  Z'_{ij}(\bm q, t) &= Z_{ij}(\bm q, t) 
  + e^{A(\bm q, t)} \psi_{ij}(\tilde{\bm k})\Delta t\\
  &\quad+ \underbrace{\left[\frac{\tilde{k}_i \tilde{k}_l}{\tilde{k}^2}Z_{mj}(\bm q, t)
  +  \frac{\tilde{k}_j \tilde{k}_l}{\tilde{k}^2} Z_{im}(\bm q, t) 
  + \gamma P_{il}(\tilde{\bm k}) Z_{mj}(\bm q, t)
  + \gamma P_{jl}(\tilde{\bm k})Z_{im}(\bm q, t)\right]}_{\Gamma_{ij}^{lm}(\bm q, t)} \overline{B_{lm}}(t) \sqrt{\Delta t} \,. \nonumber
\end{align}
Then the final step reads
\begin{align}
  Z_{ij}(\bm q, t+\Delta t) &= Z_{ij}(\bm q, t)
  + \frac12 \left[\Gamma_{ij}^{lm}(\bm q, t) + {\Gamma_{ij}^{lm}}'(\bm q, t)\right] \overline{B_{lm}}(t) \sqrt{\Delta t}
  + \frac{1}{2} \left[e^{A(\bm q, t)} \psi_{ij}(\tilde{\bm k})
  + e^{A'(\bm q, t)} \psi_{ij}(\tilde{\bm k}')\right] \Delta t
\end{align}
where ${\Gamma_{ij}^{lm}}'(\bm q, t)$ is the same as $\Gamma_{ij}^{lm}(\bm q, t)$ but with $\tilde{\bm k}(\bm q, t)$ replaced by $\tilde{\bm k}'(\bm q, t)$ and $Z_{ij}(\bm q, t)$ replaced by $Z_{ij}'(\bm q, t)$.

Finally, we write this as a scheme for $\tilde{\phi}^{\bm v}_{ij}(\bm q, t)$. The intermediate step reads
\begin{align}
  \tilde{\phi}_{ij}^{\bm v\,\prime}(\bm q, t) &= e^{-A'(\bm q, t)} Z_{ij}'(\bm q, t) \\
  &= e^{-\Delta A'(\bm q, t)}  \tilde{\phi}^{\bm v}_{ij}(\bm q, t)
  + e^{-\Delta A'(\bm q, t)} \psi_{ij}(\tilde{\bm k})\Delta t\\
  &\quad+ e^{-\Delta A'(\bm q, t)} \underbrace{\left[\frac{\tilde{k}_i \tilde{k}_l}{\tilde{k}^2}\tilde{\phi}^{\bm v}_{mj}(\bm q, t)
  +  \frac{\tilde{k}_j \tilde{k}_l}{\tilde{k}^2} \tilde{\phi}^{\bm v}_{im}(\bm q, t) 
  + \gamma P_{il}(\tilde{\bm k}) \tilde{\phi}^{\bm v}_{mj}(\bm q, t)
  + \gamma P_{jl}(\tilde{\bm k})\tilde{\phi}^{\bm v}_{im}(\bm q, t)\right]}_{\Lambda_{ij}^{lm}(\bm q, t)} \overline{B_{lm}}(t) \sqrt{\Delta t} \nonumber
\end{align}
and the final step reads
\begin{align}
  \tilde{\phi}_{ij}^{\bm v}(\bm q, t + \Delta t) &= e^{-A(\bm q, t + \Delta t)} Z_{ij}(\bm q, t + \Delta t) \\
  &= e^{-\Delta A(\bm q, t)}\tilde{\phi}^{\bm v}_{ij}(\bm q, t)
  + \frac{1}{2} e^{-\Delta A(\bm q, t)}\left[ \psi_{ij}(\tilde{\bm k})
  + e^{\Delta A'(\bm q, t)} \psi_{ij}(\tilde{\bm k}')\right] \Delta t \\
  &\quad+ \frac12 e^{-\Delta A(\bm q, t)} \left[\Lambda_{ij}^{lm}(\bm q, t) + e^{\Delta A'(\bm q, t)} {\Lambda_{ij}^{lm}}'(\bm q, t)\right] \overline{B_{lm}}(t) \sqrt{\Delta t} \nonumber
\end{align}
where, again, ${\Lambda_{ij}^{lm}}'(\bm q, t)$ is the same as ${\Lambda_{ij}^{lm}}(\bm q, t)$ but with $\tilde{\bm k}(\bm q, t)$ replaced by $\tilde{\bm k}'(\bm q, t)$ and $\tilde{\phi}^{\bm v}_{ij}(\bm q, t)$ replaced by $\tilde{\phi}^{\bm v\,\prime}_{ij}(\bm q, t)$.

\section{Numerical results for $\gamma = 0.5$}
\label{sec:numerical_results_small_gamma}
\begin{figure}[tb]
  \includegraphics{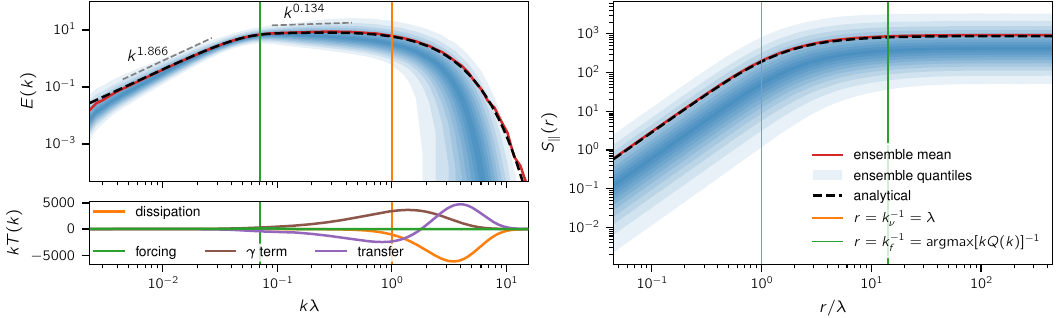}
  \caption{Energy spectrum and structure function of supplemental simulation at $\gamma = 0.5$.
  Top left: Mean energy spectrum of the ensemble simulations (red, solid) compared to the analytical solution computed from~\eqref{eq:complete_greens_function} (black, dashed). The blue, shaded area indicates the distribution of the energy spectrum across snapshots of the ensemble of simulations in the stationary state, each level set corresponding to quantiles incremented by $5\%$.
  Bottom left: Terms in the spectral energy budget equation~\eqref{app_eq:spectral_energy_budget_flux_version} (represented symbolically by $T(k)$), computed for the analytical solution. The brown line represents the energy injecting part of the $\gamma$-dependent terms in~\eqref{app_eq:spectral_energy_budget_flux_version}, which is $\frac{\beta}{4} \gamma (7 \gamma+3) E(k,t)$. The remaining $\gamma$-dependent term is contained in the transfer (purple). They are multiplied by $k$ so that the area below the curve corresponds to a rate of energy injection/dissipation in the semi-logarithmic plot. 
  Right: Mean second-order longitudinal structure function. Colors and line styles are consistent with the top left panel.}
  \label{fig:smallgamma_spec1d}
\end{figure}

\begin{table}[h]
  \caption{Parameters of the supplemental ensemble simulations.
    \label{tab:supp_sim_params}}
  \begin{ruledtabular}
  \begin{tabular}{llllllllll}
  grid size & \# members & $\beta$ & $\nu$ & $\gamma$ & $Q(k)$ & $q_0$ & $\Delta g$ & $\Delta t$ & $n_\mathrm{remap}$ \\
  $256^3$ & 40 & 16 & $10^{-3}$ & 0.5 & $2\pi k^4 \exp(-(k/2)^2)$ & 0.1 & 0.075 & $5\times 10^{-4}$ & 100
  \end{tabular}
  \end{ruledtabular}
\end{table}
As an example for convergent dynamics of the model with $\gamma \neq 0$, we conducted a supplemental set of simulations of the model at $\gamma=0.5$, using the numerical scheme described in Appendix~\ref{sec:numerics_general_gamma}. According to Appendix~\ref{sec:spectral_budget_general_gamma}, we expect convergence to a stationary solution for $|\gamma| < \sqrt{3/8} \approx 0.612$. The parameters of the simulations are listed in table~\ref{tab:supp_sim_params}. The resulting energy spectrum and structure function are shown in figure~\ref{fig:smallgamma_spec1d}.
The case $\gamma = 0.5$ features two positive power-law exponents for the energy spectrum, which are $\xi_1 \approx 0.134$ and $\xi_2 \approx 1.866$ according to~\eqref{eq:stationary_power_law_solutions_exponents}. Figure~\ref{fig:smallgamma_spec1d} (upper left) shows that these are both realized for wavenumbers beyond and below the forcing range, respectively. Analogous to figure~\ref{fig:spec1d}, we compare the mean energy spectrum from the ensemble simulations (red, solid) to the analytical solution computed from~\eqref{eq:complete_greens_function} (black, dashed). Despite the positive power-law exponents, the solution does not diverge due to the viscous cut-off. Shades of blue indicate the distribution of the energy spectrum across different realizations (snapshots) of the field. The lower left panel of figure~\ref{fig:smallgamma_spec1d} shows the terms of the spectral energy budget equation~\eqref{app_eq:spectral_energy_budget_flux_version}. In this case, most of the energy is generated through the $\gamma$ term $\frac{\beta}{4} \gamma (7 \gamma+3) E(k,t)$, and the forcing is almost negligible in comparison. Note that the transfer term also depends on $\gamma$. The right panel shows the second-order structure function, which looks very comparable to the $\gamma=0$ case, albeit with stronger fluctuations.

\bibliography{../../refs.bib}

\end{document}